\documentclass[journal,12pt,onecolumn,letterpaper]{IEEEtran}
\usepackage{arxiv}
\usepackage{geometry}
\usepackage{times}
\usepackage{cite}
\usepackage{url}
\usepackage{graphicx}
\usepackage{lscape}
\usepackage{subfigure}
\usepackage{rotating}
\usepackage{rotfloat}
\usepackage{xcolor}
\usepackage{amsmath}
\usepackage{amssymb}
\usepackage[linesnumbered,ruled,vlined]{algorithm2e}
\usepackage{pseudocode}
\usepackage{array}
\usepackage[english]{babel}
\usepackage{gensymb}
\usepackage{textcomp}
\usepackage{placeins}
\usepackage{balance}
\usepackage{booktabs}

\title{SimplyMime: A Control at Our Fingertips}

\author{%
    Sibi Chakkaravarthy Sethuraman\\
    Centre of Excellence, Artificial Intelligence \& Robotics (AIR),\\
    School of Computer Science and Engineering\\
    VIT-AP University, India \\
    \texttt{sb.sibi@gmail.com} \\
\And
    Gaurav Reddy Tadkapally \\
    Centre of Excellence, Artificial Intelligence \& Robotics (AIR),\\
    School of Computer Science and Engineering\\
    VIT-AP University, India \\
    \texttt{gauravreddy008@gmail.com}
\And
    Athresh Kiran \\
    Senior Software Developer\\
    Parallel Reality: AI-based Health Tech \\
    United Kingdom \\	
  \texttt { athresh.kiran@gmail.com}
\And
    Saraju P. Mohanty\\
    Department of Computer Science and Engineering,\\
    University of North Texas, \\
    TX 76207, USA, \\
    \texttt{saraju.mohanty@unt.edu}\\
\And
    Anitha Subramanian \\
    Centre of Excellence, Artificial Intelligence \& Robotics (AIR),\\
    School of Computer Science and Engineering\\
    VIT-AP University, India \\
\texttt{anithachubbu@gmail.com}
}
\begin{document}

   

\maketitle

\begin{abstract}
The utilization of consumer electronics, such as televisions, set-top boxes, home theaters, and air conditioners, has become increasingly prevalent in modern society as technology continues to evolve. As new devices enter our homes each year, the accumulation of multiple infrared remote controls to operate them not only results in a waste of energy and resources, but also creates a cumbersome and cluttered environment for the user. This paper presents a novel system, named SimplyMime, which aims to eliminate the need for multiple remote controls for consumer electronics and provide the user with intuitive control without the need for additional devices.

SimplyMime leverages a dynamic hand gesture recognition architecture, incorporating Artificial Intelligence and Human-Computer Interaction, to create a sophisticated system that enables users to interact with a vast majority of consumer electronics with ease. Additionally, SimplyMime has a security aspect where it can verify and authenticate the user utilising the palmprint, which ensures that only authorized users can control the devices. The performance of the proposed method for detecting and recognizing gestures in a stream of motion was thoroughly tested and validated using multiple benchmark datasets, resulting in commendable accuracy levels.

One of the distinct advantages of the proposed method is its minimal computational power requirements, making it highly adaptable and reliable in a wide range of circumstances. The paper proposes incorporating this technology into all consumer electronic devices that currently require a secondary remote for operation, thus promoting a more efficient and sustainable living environment.
\end{abstract}


\enlargethispage{10pt}
\section{Introduction}

Smart electronics, such as televisions, air conditioners, speakers, and ceiling fans, have become ubiquitous in modern society. Thus driving the framework of smart cities and smart villages which are intended to operate optimally with limited resources while best utilizing the available resources to improve quality of life \cite{Joshi_iGLU2, Rachakonda_Stress-Lysis}. With the proliferation of inexpensive and readily available devices, most households have multiple electronic devices that require remote controls for operation and interaction \cite{andrae_life_2010}. As the number of devices increases each year, so too does the number of remote controls needed to operate them. This traditional method of interaction, while widespread and commonly used, is not without its flaws. The use of multiple remote controls not only wastes resources and increases the use of plastic, but also makes it difficult to locate and operate the correct remote for a given device \cite{issa_usability_2022}. 

The first remote-control devices were introduced in the 1950s, but it wasn't until the 1970s that infrared-based remotes began to appear on the market \cite{oudah_hand_2020}. Despite some advancements in human-computer interaction (HCI) technology, such as keyboard and mouse alternatives and devices that use Bluetooth or WiFi communication, these new approaches have not been able to fully replace traditional remote controls due to a lack of seamless functionality. However, recent advances in HCI technology, such as voice commands, mimics, and gestures used in devices such as tablets, smartphones, and smart homes, have shown that there is still potential for improvement in the field \cite{seiter_:_2013}.

SimplyMime is a hand gesture-based control system for consumer electronics that addresses the shortcomings of traditional remote controls. Hand tracking is a natural and intuitive mechanism for identifying hand movements, and it has been studied for a long time. Skeleton-based hand tracking, due to its resistance to various background conditions, has proven to be a popular choice for this type of technology \cite{sodhi_analysis_2007}. SimplyMime offers a lightweight solution that can track hands and provide real-time output, enabling it to be incorporated into a compact system that can be installed in any electronic appliance. Furthermore, traditional remote controls lack security features. To further enhance security, SimplyMime incorporates facial recognition and detection to authenticate and validate the user's identity. This added security measure ensures that only authorized users are able to control the devices, providing an additional layer of protection for users. 

\begin{figure*}[ht]
\centering
\includegraphics[width=\textwidth]{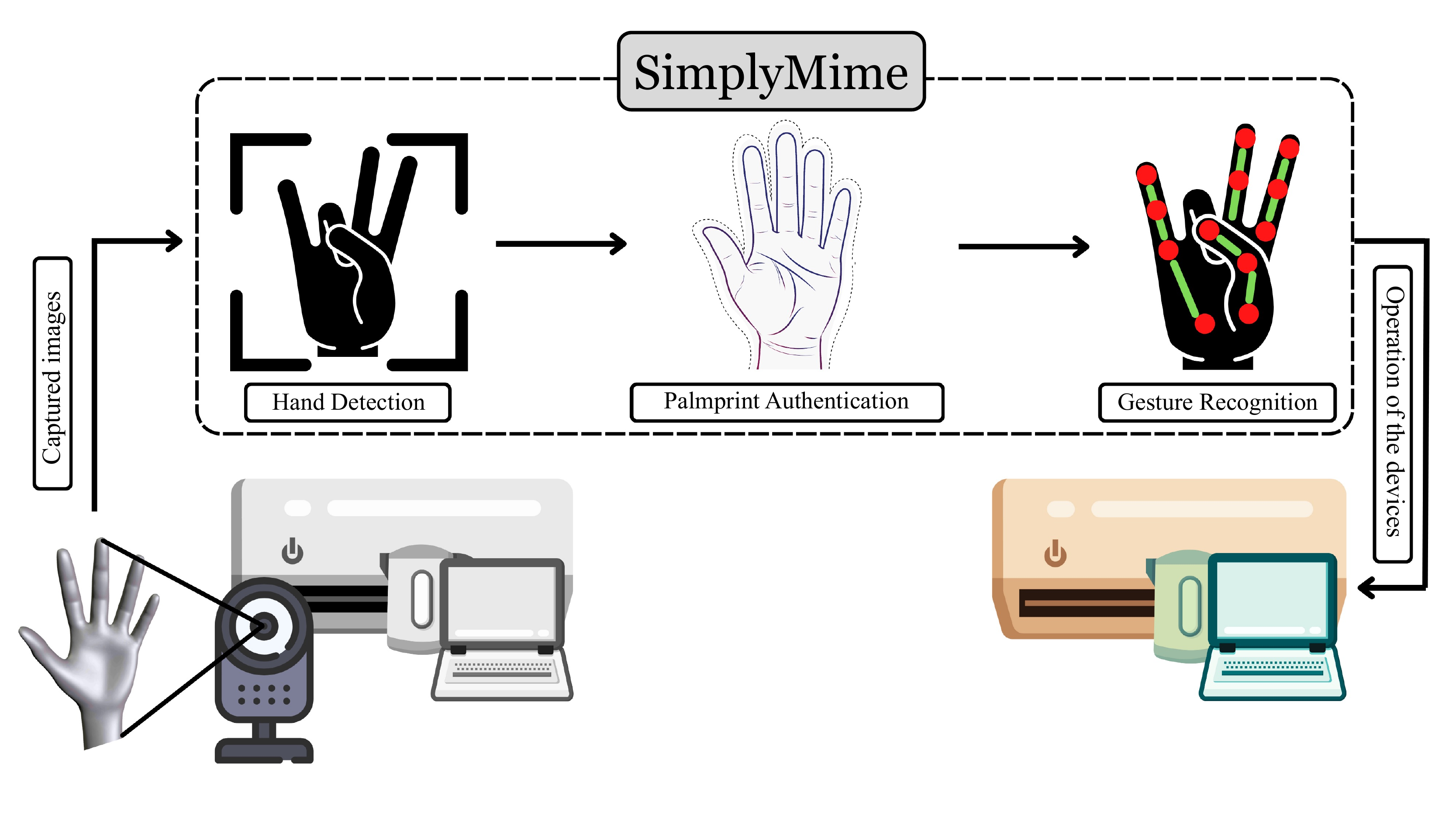}
\caption{Diagrammatic representation of the architectural working of SimplyMime}
\label{FIG:Simplified_Architecture_SimplyMime}
\end{figure*}

SimplyMime harnesses the power of hand detection technology to recognize various user gestures, as outlined in Section \ref{Sec:Proposed_Simplymime}. This allows for the execution of specific actions, such as controlling embedded devices, such as turning them ON/OFF, adjusting speeds and volumes, among others. It offers an efficient and intuitive means of interacting with consumer electronics.

The main objectives of SimplyMime are the following:

\begin{itemize}
    \item \textbf{User-centric Design}: As discussed in Section \ref{Sec:Related_Research}, various technologies have been proposed to control different devices, but few have given consideration to the user's convenience. SimplyMime prioritizes ease of use for the end user.
    \item \textbf{Seamless Integration}: The SimplyMime system is designed to be integrated seamlessly into any device without altering its internal architecture, making it easy for users to migrate from traditional remote controls to this new technology.
    \item \textbf{Sustainable Solutions}: SimplyMime offers a one-stop solution for controlling multiple devices, reducing the accumulation of e-waste such as remote controls and other controllers.
    \item \textbf{Advanced Security}: SimplyMime includes a security authentication component that verifies and authenticates the user before granting access, providing an added layer of security, unlike traditional remote controls.
\end{itemize}

Rest of the paper is organized as follows: Section \ref{Sec:Novel_Contributions} outlines novel contributions of the current paper. Section \ref{Sec:Related_Research} discusses about the existing related research. Section \ref{Sec:Proposed_Simplymime} provides the system level architecture of the proposed SimplyMime and outlines the novel methods proposed for hand detection, hand landmark detection and Palm print authentication. Section \ref{Sec:Experimental_validation} is used to validate the results of the proposed SimplyMime system and also provides a comparative analysis of SimplyMime against the state of the art solutions. Finally, the paper concludes in Section \ref{Sec:Conclusion}.

\section{Novel Contributions of the Current Paper}
\label{Sec:Novel_Contributions}

\subsubsection{Problem Addressed}
The traditional remote controls require line of sight and precise pointing, hindering the user's freedom of movement and reducing the overall comfort level. As discussed earlier, even the recent contributions fail to develop a solution which eliminated the requirement of additional device completely, and provides a better intuitive experience to the user, resulting in unreliable and frustrating user experiences. 

\subsubsection{Solution Proposed by SimplyMime}
SimplyMime is a state-of-the-art system that employs advanced hand gesture recognition techniques, leveraging the latest advancements in Artificial Intelligence and Human-Computer Interaction, to create a highly efficient and intuitive control framework for consumer electronics. The system utilizes a camera to capture the user's hand gestures, which are then translated into commands for electronic devices. The dynamic architecture of SimplyMime enables it to adapt to various hand gestures and provide a seamless user experience. The architecture is equipped with a palm print authentication module that ensures the device is only accessible to authorized users.

The SimplyMime system is designed to eliminate the need for multiple remote controls, providing users with a streamlined and organized living space. By removing the need for additional devices, SimplyMime offers a more sustainable solution that saves energy and resources. Moreover, the intuitive nature of the system provides users with greater comfort and convenience, as they no longer have to search for and operate multiple remote controls. Additionally, the dynamic architecture of SimplyMime enables it to adapt to new gestures, ensuring that it can keep up with the changing needs of users.

\subsubsection{Significance of the Proposed Solution}

Our proposed solution, SimplyMime, offers several advantages over traditional remote control systems. First, it eliminates the need for multiple controllers, leading to a cleaner, more organized living space. Second, the system offers a more immersive and intuitive experience for users, making it easier to interact with electronic devices. Moreover, our pipeline incorporates a security feature that is often overlooked in previous solutions, by enabling verification and authentication of the user through the use of palmprint recognition. This added security measure ensures that only authorized users are able to control the devices, providing an additional layer of protection for users. We believe that SimplyMime's innovative approach to remote control systems has the potential to revolutionize the way users interact with their consumer electronics. Its minimal computational requirements make it highly adaptable and reliable in a wide range of circumstances, making it an ideal solution for households and businesses alike.

\section{Prior Related Research}
\label{Sec:Related_Research}

Hand gesture recognition technology has been a subject of study for several years, with developments in different fields, including human-computer interaction, gaming, augmented reality, and assistive technology for those with disabilities. The hand, in particular, is used extensively for gesturing compared to other body parts because it is a natural means of human communication and therefore the most appropriate tool for HCI \cite{sodhi_analysis_2007}. And evidently, controlling electronic devices with our hands is a natural and highly intuitive method of interaction. The use of hand gesture recognition to operate consumer electronics such as televisions and home appliances is also gaining traction \cite{rautaray_vision_2015}. This application area has the potential to revolutionise how we interact with electronic devices in our homes by enabling simple hand gestures to control them. However the efforts to develop a reliable and robust hand gesture recognition system for this application ares is still lacking.

Variability in how individuals make gestures is one of the greatest obstacles to hand gesture recognition. People can perform the same gesture differently based on their age, gender, cultural background, and physical capabilities. Researchers have developed various techniques for capturing and analysing hand gestures, including computer vision, machine learning, and sensor-based approaches \cite{hasan_human_2012}.

\subsection{Sensor-based Methods}
The majority of existing methods can be divided into two categories, with sensor-based systems constituting the first. Utilizing specialised sensors to capture the movement and orientation of the hand, sensor-based methods for hand gesture recognition capture the hand's motion and the orientation. These sensors may include accelerometers, gyroscopes, magnetometers, and other types of sensors that can measure the hand's movement and position \cite{guo_human-machine_2021}. Sensor-based methods have a number of advantages over other methods, including the ability to capture precise information about the hand's movement and position and the capacity to function in environments with poor lighting or visibility. In a recent study, the authors employed the utilization of inertial sensors, specifically accelerometers and gyroscopes, in conjunction with a multilayered perceptron classifier to recognize hand gestures \cite{teachasrisaksakul_hand_2018}. Adopting a similar methodology, other researchers have developed a finger-worn ring device that captures acceleration data, paired with an LSTM model for the classification of hand gestures \cite{coffen_tinydl:_2021, nguyen-trong_gesture_2021, yuan_hand_2021}. Likewise, Inviz device incorporates textile-based flexible capacitive sensor arrays for motion detection \cite{singh_inviz:_2015}. This approach has been utilized to develop systems that cater to individuals with paralysis, paresis, weakness, and restricted range of motion. Other notable developments include the WristCam \cite{chen_wristcam:_2019}, a wrist-worn camera sensor that estimates and recognizes hand trajectories, as well as the EchoFlex \cite{mcintosh_echoflex:_2017}, a wearable ultrasonic device that tracks hand movements through a novel method. The research makes use of the muscle flexor data collected by the proposed device to develop an algorithm for gesture recognition that combines image processing and neural networks. 

Sensor-based systems provide precise information about the hand's movement and position and can function in environments with poor lighting or visibility. However, the requirement for the user to wear additional devices restricts the versatility and scalability of these systems. In many instances, faulty sensor calibration can result in errors in gesture recognition, resulting in abysmal performance.

\subsection{Vision-based Methods}
The integration of computer vision and human-computer interaction has enabled the development of innovative systems aimed at addressing limitations and providing users with a more immersive and efficient experience when interacting with machines. One of the initial approaches in the development of vision-based sensors was the utilization of a colored glove \cite{wang_real-time_2009}. This method required the user to wear a colored glove, which enabled the system to employ a nearest-neighbor approach for tracking hand movements at interactive rates. Several studies that employed the use of the Kinect sensor utilized color and image depth data to establish a hand model for the analysis of tracked hand gestures. The hand gesture tracking was accomplished through the implementation of a Kalman filter \cite{feng_fusing_2014}. However, it was observed that the gesture recognition accuracy was relatively low. Similar research utilized RGB-D cameras to extract hand location data through the use of in-depth skeleton-joint information from images \cite{xu_robust_2022, tran_real-time_2021}. Furthermore, similar to the architecture of SimplyMime, a few works employed neural networks to infer real-time hand landmarks, which were used to establish a skeletal structure of the hand gesture \cite{shin_skeleton-based_2020, caputo_shrec_2021}.

Computer vision-based systems are non-intrusive and more versatile than sensor-based systems, but they can be affected by poor lighting or visibility. Moreover, recognizing gestures from complex backgrounds is still a challenging task. As a reason, SimplyMime focuses on developing robust and reliable hand gesture recognition systems that can overcome these challenges and enable natural and intuitive interactions with electronic devices.

\begin{table*}[t]
\centering
\caption{Comparison of existing hand gesture recognition models}
\label{tab:handrec_results}
\resizebox{\textwidth}{!}{%
\begin{tabular}{@{}llllll@{}}
\hline 
Research                                                                                                                             & \begin{tabular}[c]{@{}l@{}}Methodology\\ Employed\end{tabular}                      & Findings and Outcome                                                                                                                                                                                        & \begin{tabular}[c]{@{}l@{}}Requirement of\\ additional device\end{tabular}            & \begin{tabular}[c]{@{}l@{}}Requirement \\of Calibration\end{tabular} & \begin{tabular}[c]{@{}l@{}}Security\\ Module\end{tabular}                  \\ \midrule
\hline
\begin{tabular}[c]{@{}l@{}}Teachasrisaksakul et al.\\ (2018) \cite{teachasrisaksakul_hand_2018}  \end{tabular} & Inertial Sensors                                                                    & \begin{tabular}[c]{@{}l@{}}- Achieved an accuracy of 98.33\%.\\- However, the performance will be impacted\\ by external factors, such as the user's body\\ movements or the environmental conditions \end{tabular}                                                         & \begin{tabular}[c]{@{}l@{}}Accelerometer and\\ Gyroscope\end{tabular}                 & Yes                                                                  & None                                                                       \\ \\
\begin{tabular}[c]{@{}l@{}}TinyDL (2021) \\ \cite{coffen_tinydl:_2021} \end{tabular}                                & Inertial Sensors                                                                    & \begin{tabular}[c]{@{}l@{}}- Utilized a built-in LSTM model that leveraged\\ data from a finger-worn ring.\\- The performance will be impacted by factors \\such as sensor placement, variations among users, \\and changes in hand gesture execution over time.\end{tabular} & Finger-worn device                                                                    & Yes                                                                  & None                                                                       \\ \\
\begin{tabular}[c]{@{}l@{}}Inviz (2015)\\ \cite{singh_inviz:_2015}\end{tabular}                                   & \begin{tabular}[c]{@{}l@{}}Textile-based\\ capacitive arrays\end{tabular}           & \begin{tabular}[c]{@{}l@{}}- Relies on textile-based capacitive arrays built into\\ clothing.\\- The production of wearable textile capacitive\\ sensor arrays could increase the cost and \\complexity of the system.\end{tabular}                                           & \begin{tabular}[c]{@{}l@{}}Wearable textile\\ capacitive sensor\\ Arrays\end{tabular} & Yes                                                                  & None                                                                       \\ \\
\begin{tabular}[c]{@{}l@{}}EchoFlex\\ (2017) \cite{mcintosh_echoflex:_2017} \end{tabular}          & \begin{tabular}[c]{@{}l@{}}Ultrasound\\ Imaging\end{tabular}                        & \begin{tabular}[c]{@{}l@{}}- Utilizes ultrasound sensors to capture hand\\ movement data, which is then analyzed using ML\\ algorithms. \\- Susceptible to interference from nearby objects\\ or environments with high acoustic noise\end{tabular}                           & \begin{tabular}[c]{@{}l@{}}Ultrasonographic\\ Device\end{tabular}                     & Yes                                                                  & None                                                                       \\ \\
\begin{tabular}[c]{@{}l@{}}Wang et al. (2009)\\ \cite{wang_real-time_2009}\end{tabular}                        & Image Processing                                                                    & \begin{tabular}[c]{@{}l@{}}- Employs colour segmentation and a tracking\\ algorithm is applied to estimate the hand pose and\\ motion.\\- Recognition accuracy would be affected by\\ factors such as lighting conditions, colour \\variations, and occlusions.\end{tabular}  & Coloured Glove                                                                        & No                                                                   & None                                                                       \\ \\
\begin{tabular}[c]{@{}l@{}}Feng et al. (2014)\\ \cite{feng_fusing_2014} \end{tabular}                              & Kinect Sensors                                                                      & \begin{tabular}[c]{@{}l@{}}- Demonstrates the feasibility of combining various\\ types of sensors for human motion tracking.\\- Requires a high-performance computing device\\ to achieve real-time performance\end{tabular}                                                 & \begin{tabular}[c]{@{}l@{}}Microsoft Kinect\\ Sensor\end{tabular}                     & Yes                                                                  & None                                                                       \\ \\
\begin{tabular}[c]{@{}l@{}}SHREC (2021)\\ \cite{caputo_shrec_2021} \end{tabular}                                   & \begin{tabular}[c]{@{}l@{}}Skeleton-based\\ hand gesture\\ Recognition\end{tabular} & \begin{tabular}[c]{@{}l@{}}-  Achieved high accuracy in recognizing hand\\ gestures in various real-world scenarios\end{tabular}                                                                                                                                               & None                                                                                  & No                                                                   & None                                                                       \\ \\
\begin{tabular}[c]{@{}l@{}}\textbf{SimplyMime} \\ \textbf{(Current Paper)}\end{tabular}                                                                & \begin{tabular}[c]{@{}l@{}}CNN based\\ Skeletal Pose\\ Estimation\end{tabular}      & \begin{tabular}[c]{@{}l@{}}- Delivers intuitive control without additional \\devices\\- Incorporates palmprint authentication to verify\\ and authenticate users\end{tabular}                                                                                                 & None                                                                                  & No                                                                   & \begin{tabular}[c]{@{}l@{}}Palmprint based\\ Authentication\end{tabular} \\ \bottomrule
\end{tabular}%
}
\end{table*}

\enlargethispage{10pt}
\section{SimplyMime: The Proposed Smart Remote Control}
\label{Sec:Proposed_Simplymime}

The proliferation of consumer electronics has resulted in the widespread use of traditional remote controls as the primary means of interaction. However, in order to effectively replace such a firmly established technology, an alternative solution must not only be robust, precise, and intuitive, but also possess the added benefits of user-friendliness, compatibility with older devices, and scalability \cite{rautaray_vision_2015}. Our proposed work, SimplyMime, addresses the shortcomings of existing solutions, while also maintaining the immersive experience that traditional remote controls are capable of delivering. The dynamic hand gesture recognition module, represents the most advanced, effective, and ideal replacement for traditional remote controls, providing a unique blend of state-of-the-art performance and efficiency.

The underlying architecture of the system incorporates a hand landmark assignment method, which is utilized to identify and localize key points across the hand, such as finger tips, knuckles, and wrists. These landmarks, assigned by the backend model, form the basis for the gesture recognition algorithm, which is able to identify and classify gestures based on the skeletal structure generated. The system is designed to operate in two distinct modules, which are discussed in greater detail in the following sections. The first module, the hand detection model, is responsible for identifying and isolating the human hand within an image. The second module, the gesture recognition model, processes the detected hand to generate a skeletal structure of the gesture and subsequently recognizes it. The suggested approach constitutes a noteworthy progression in the domain of hand gesture detection, showing potential as a viable substitute for conventional human-computer interaction techniques.

\subsection{Hand Detection Model}
The foundation of an effective hand gesture identification model is the accurate detection and isolation of the hand from the given image. To achieve this, SimplyMime employs state-of-the-art neural network technology for localizing the coordinates of the palm. Our model specifically utilizes the Single Shot Detector (SSD) architecture \cite{liu_ssd:_2016}. While conventional RCNN models have been widely used for object localization, they require significant computational power, making them less reliable for real-time applications \cite{girshick_rich_2014}. In contrast, our SSD algorithm achieves superior real-time performance by eliminating the need for bounding box proposals and the subsequent feature re-sampling stage. This allows for a more efficient and effective approach to hand detection, which is a crucial component of an accurate gesture identification system.

\begin{figure} [htbp]
\centering
	\includegraphics[width=0.8\textwidth]{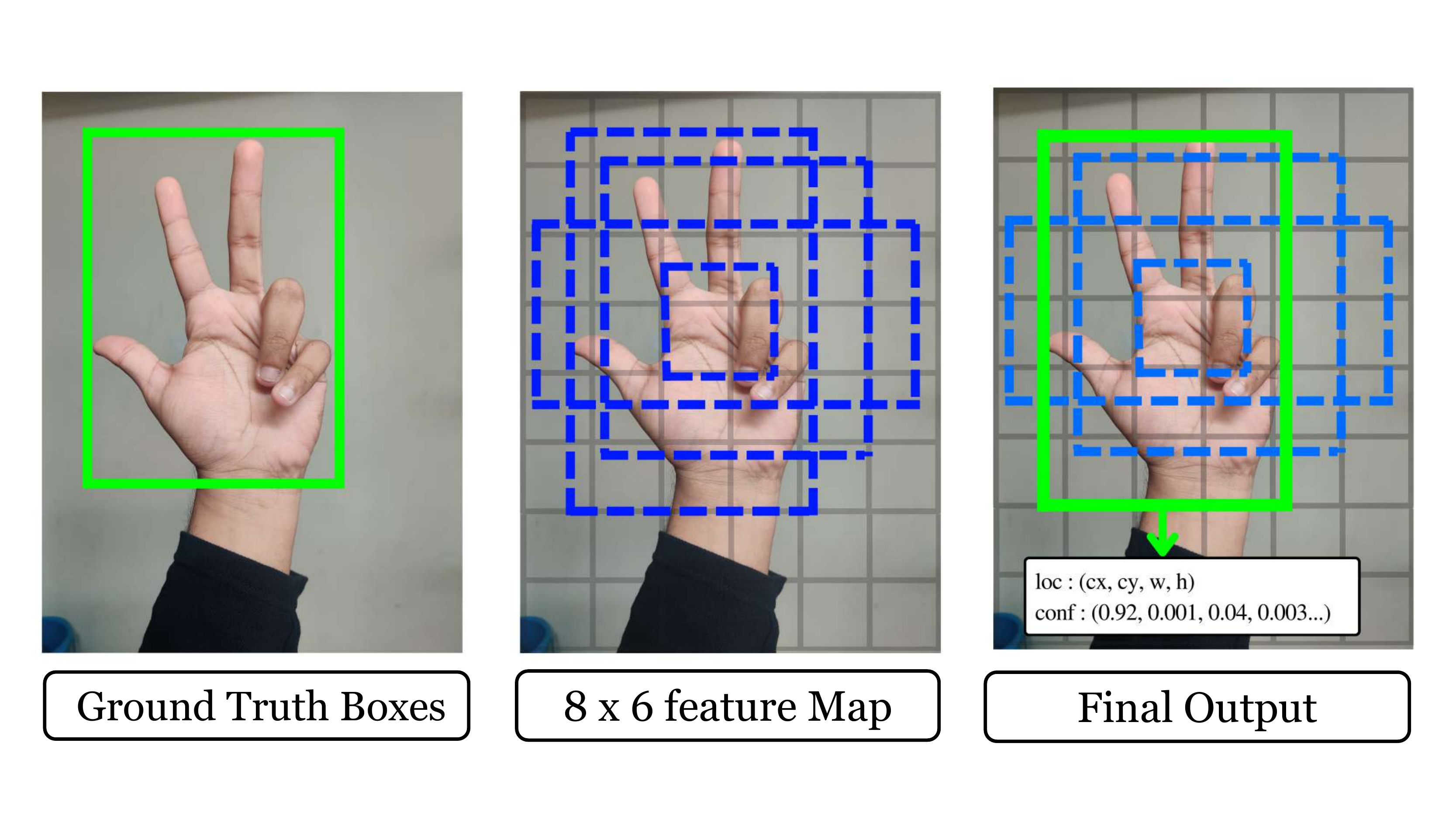}
\caption{Working overview of the Hand detection model}
\label{FIG:anchor_boxes}
\end{figure}

\begin{figure} [htbp]
\centering
	\includegraphics[width=0.8\textwidth]{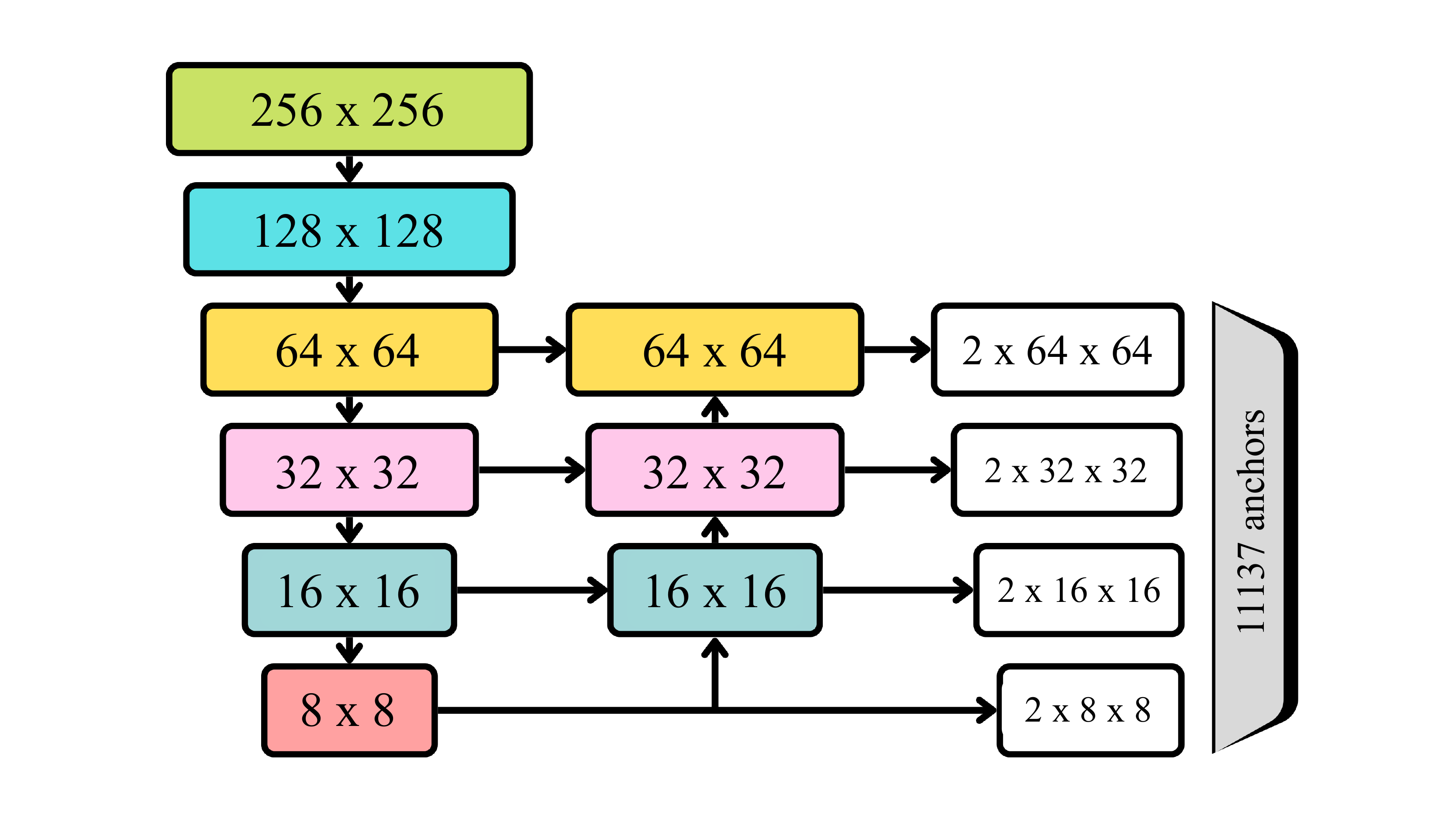}
\caption{Comprehensive view of the Gesture recognition module's architecture}
\label{FIG:handdet_arch}
\end{figure}

\begin{figure} [htbp]
\centering
	\includegraphics[width=0.7\textwidth]{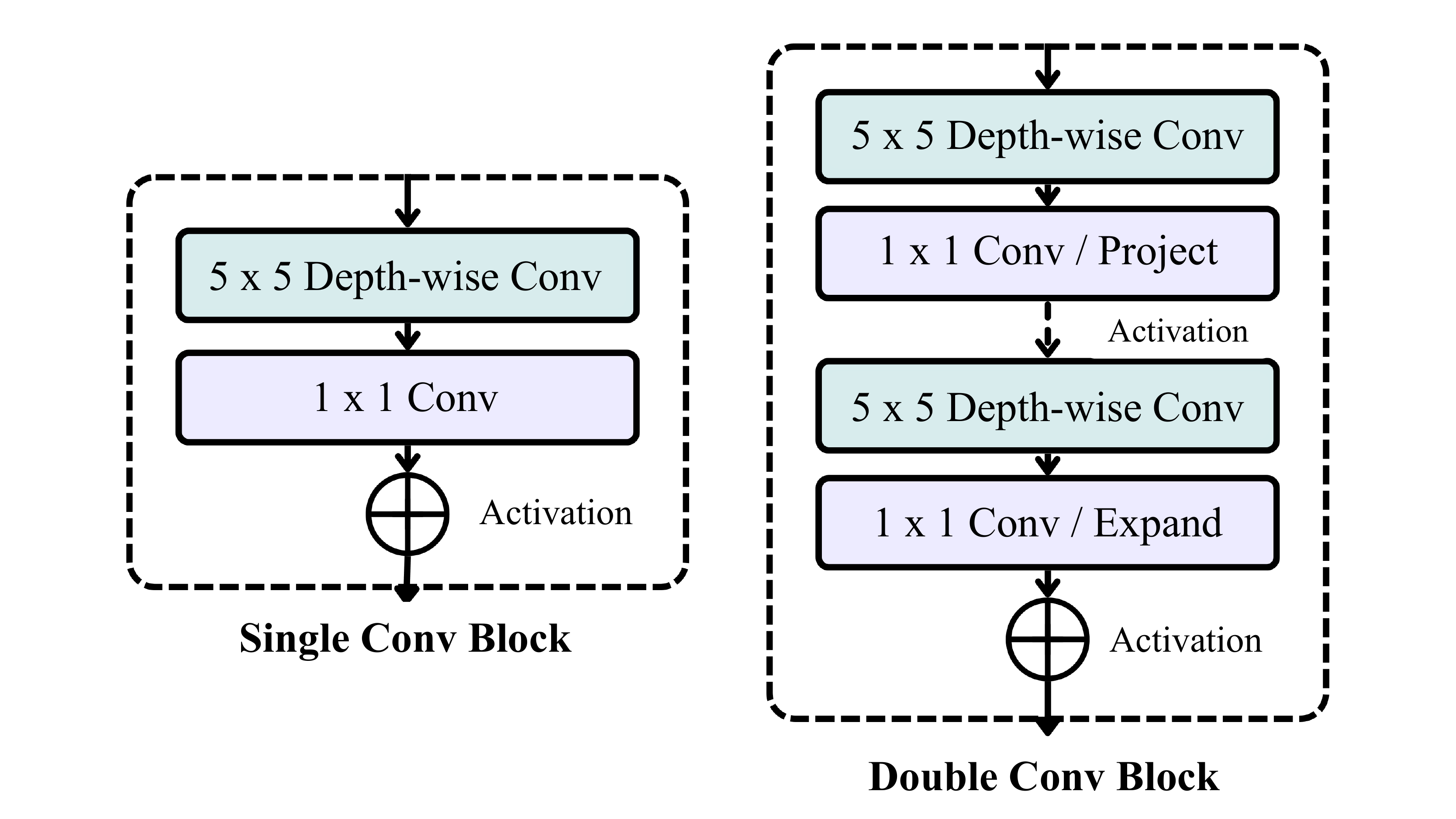}
\caption{Inner structure of the devised Convolution blocks}
\label{FIG:convblocks}
\end{figure}

The hand detection model employed in SimplyMime utilizes a unique approach in which it is initially trained on human palms rather than the entire hand. This approach allows the network to more easily perceive patterns and generate bounding boxes, as the palms and fists are more rigid objects when compared to a human hand with articulated fingers. In the task of detecting human hands, the last layer of the model creates substantially smaller anchor boxes, as human hands are smaller objects and thus require smaller anchor boxes. The architectural network of SimplyMime's detection model is illustrated in Figure \ref{FIG:handdet_arch}. The feature extraction network takes an input RGB image of 224x224px and is followed by a series of convolutional layers, referred to as ConvBlocks, which serve as the bottleneck for the higher abstraction level layers. Essentially, the image is passed through 5 single and 6 double ConvBlocks. The introduced ConvBlocks consist of a series of convolutional layers, followed by a depth-wise and a point-wise convolutional layer. The detailed architecture of our Convblocks is depicted in the Figure \ref{FIG:convblocks}. Furthermore, our network outperformed a popular light-weight model, MobileNetV2-SSD \cite{howard_mobilenets:_2017}, in terms of both accuracy and inference speed.

The detection architecture of SimplyMime is designed to be highly robust and accurate, and this is achieved through the use of a diverse set of training data. To ensure that the model is able to generalize well to a wide range of scenarios, it is initially trained on three different sources of data. The first of these is an in-the-wild dataset, which comprises a diverse set of images captured in various geographical locations and under different lighting conditions, allowing the model to learn to detect hands under a wide variety of conditions. The second data source is an in-house dataset, which is specifically designed to cover all possible hand angles in a controlled environment. This dataset allows the model to learn to detect hands under consistent conditions, and the combination of these two datasets allows for a well-represented and diversified dataset. Finally, the model is further trained on a synthetic dataset to ensure that it is able to detect hands under a wide range of angles and in a variety of environments, further improving its robustness and accuracy.

\subsection{Hand Landmark Detection Model}

\begin{figure*} [htbp]
\centering
	\includegraphics[width=\textwidth]{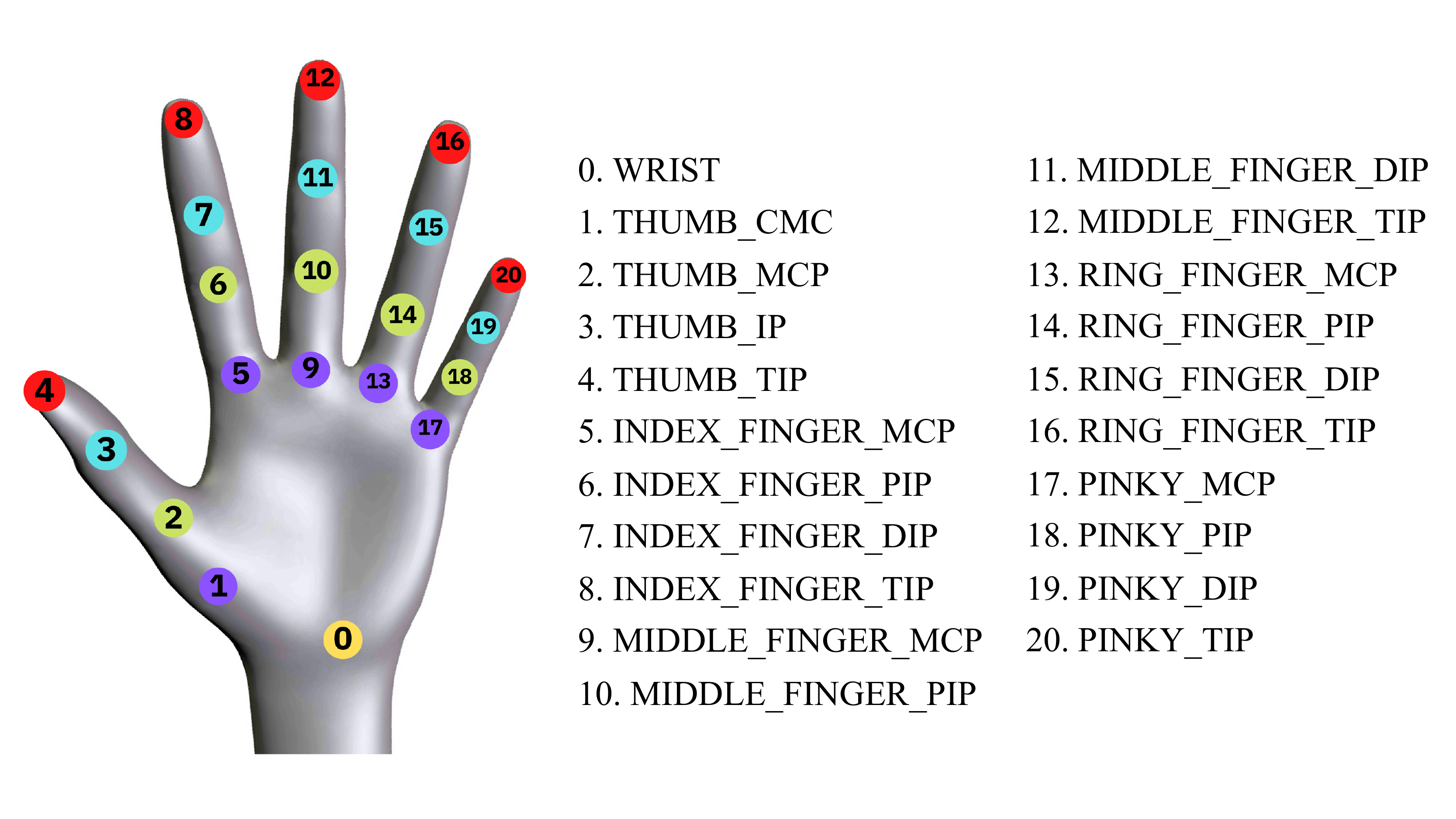}
\caption{Key-point indexes of all hand landmarks}
\label{FIG:landmarks}
\end{figure*}

Having successfully localized the hand in the image, the next step in SimplyMime's pipeline is to extract key-points from the isolated hand region. For this purpose, we employ a modified version of the Convolutional Pose Machines (CPM) network, which has been extensively used in the field of human pose estimation \cite{wei_convolutional_2016}. The CPMs provide a confidence map for each keypoint, represented as a Gaussian centered at the true position. These maps are generated based on the size of the input image patch, and the final location of each keypoint is determined by identifying the peak in the corresponding confidence map. By assigning 21 landmarks across different points across the palm, the model achieves to generate a skeletal structure of the palm.

\subsubsection*{\textbf{Dynamic Gesture Detection Algorithm}}
After extracting the keypoints from the hand region, they are transferred to the gesture detection engine for further analysis. These keypoints, which consist of 21 points, represent distinct areas of the hand and are illustrated in Figure \ref{FIG:landmarks}. They serve as the basis for identifying the gesture that the user intends to create. This processed information is then utilized to control the SimplyMime-enabled devices. Our algorithm incorporates a sophisticated technique that assigns a FingerState value to each digit. An open finger is designated as 1, while a folded finger is denoted as 0. To determine the FingerState, we leveraged the Y-axis coordinates of the metacarpophalangeal (MCP) joint and the fingertip. Nevertheless, the thumb's unique location and alignment necessitated calculating its slope to ascertain its FingerState. This approach results in more precise and accurate gesture recognition and enhances the system's overall performance.

\begin{figure*} [htbp]
\centering
	\includegraphics[width=\textwidth]{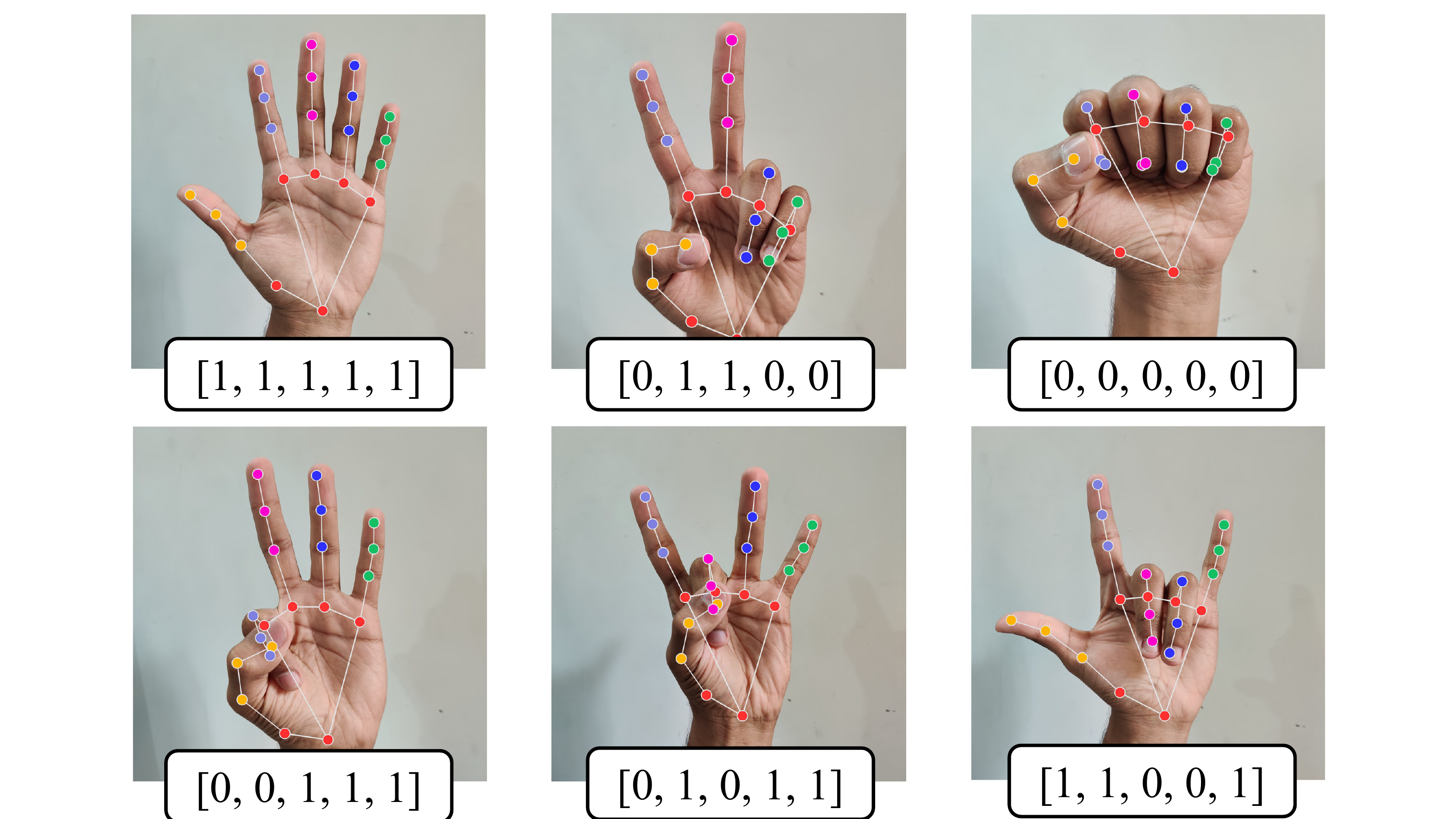}
\caption{Working results and subsequent arrays generated by the landmark detection model}
\label{FIG:landmarks_working}
\end{figure*}

The concluding stage of the process involves the recognition of gestures by analyzing the extracted keypoints from the hand region. Through the use of a posture array, each gesture can be accurately identified and differentiated from one another. These posture arrays, each unique to a particular gesture, can be designated as trigger functions to regulate a diverse range of consumer electronics. To add a dynamic element to the algorithm, the metacarpophalangeal joint (MCP) of the middle finger is employed as the focal point to align and center the camera, ensuring that the palm remains in view. Moreover, the midpoint between the tips of the thumb and index finger is used to locate the cursor, if necessary, for a specific device. Figure \ref{FIG:landmarks_working} demonstrates several images and their corresponding posture arrays.

\subsection{Palm Print Authentication Module}

The protection of security is of paramount importance in hand gesture-based systems utilized for controlling consumer electronics, as these systems are susceptible to unauthorized access and control \cite{cyril_jose_smart_2015}. Without robust security mechanisms, these systems can be easily compromised, leading to data breaches, unauthorized access to sensitive information, and potential damage to devices. Therefore, it is essential to incorporate advanced security measures during the design and development of hand gesture-based systems to mitigate these risks. The integration of strong security measures is critical for ensuring the reliability and integrity of hand gesture-based systems for consumer electronics.

SimplyMime employs advanced biometric authentication methods, specifically, PalmPrint identification. To achieve this, we utilize a Siamese Network architecture, a common approach used in similarity recognition tasks like facial recognition \cite{muller_few-shot_2022}. Our network includes two parallel feature extraction blocks that process two distinct palm images, producing a 4096-dimensional tensor for each image. By measuring the dissimilarity between the two sets of embeddings utilizing the euclidean distance function, the similarity between the two palms is verified. A pre-determined threshold is established to establish the authenticity of the user. Access to the system is restricted if the dissimilarity measure exceeds the threshold, ensuring that only authorized individuals have access to the system. This multi-factor approach ensures that only authorized individuals have access to the system, protecting against unauthorized attempts to control the consumer electronics. Furthermore, implementing these security measures guarantees the protection of personal and sensitive information, enhancing the system's reliability and integrity.

\begin{figure*} [htbp]
\centering
	\includegraphics[width=\textwidth]{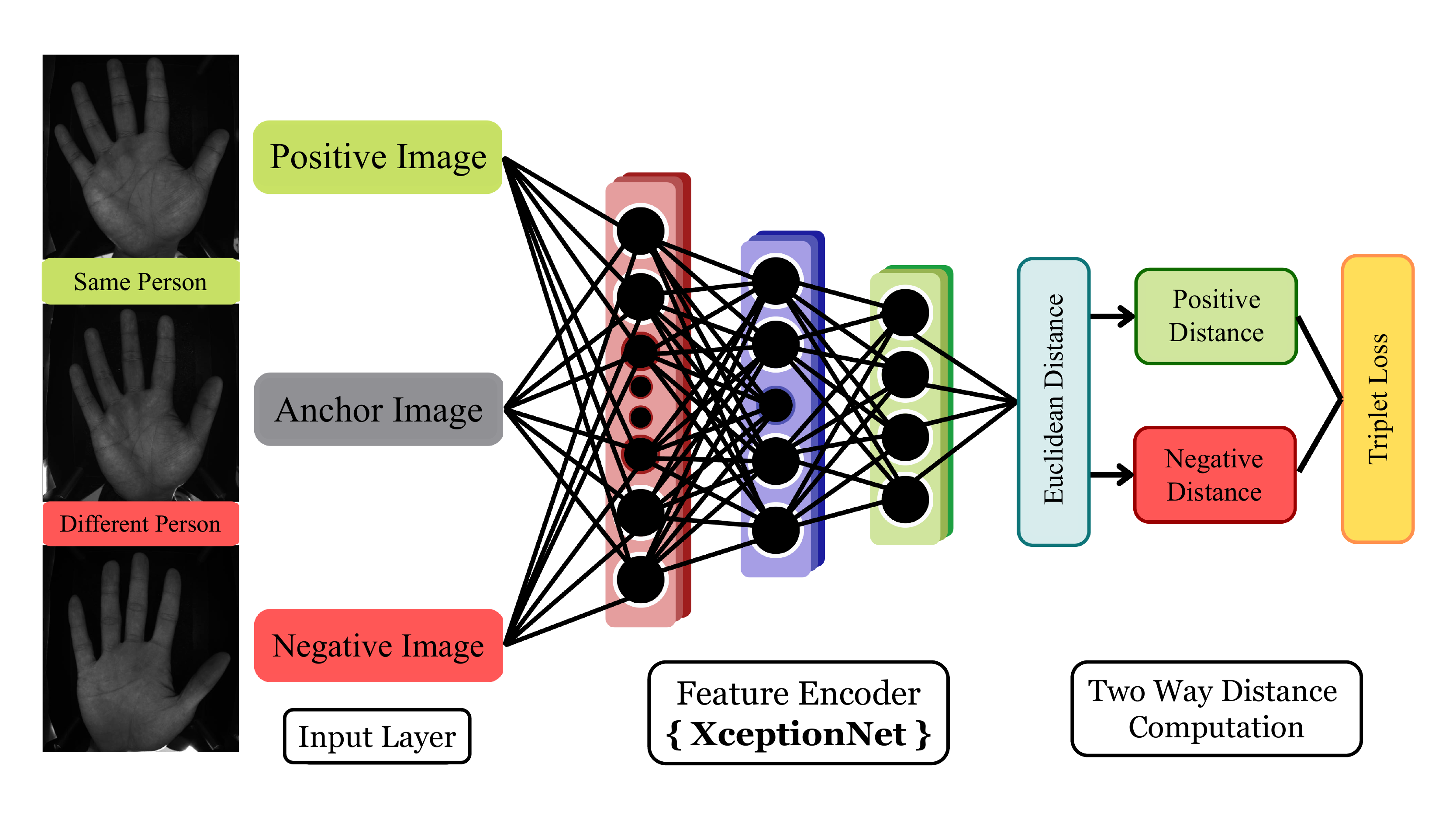}
\caption{Working pipeline of the Siamese network based authentication model}
\label{FIG:siamese_working}
\end{figure*}

During the training process, the network is presented with three input images, including an anchor image, a positive image, and a negative image. The anchor and positive images correspond to the palm print of the same individual, while the negative image represents the palm print of a different individual. The network is trained for 100 epochs using the Triplet Loss function, which is optimized using Adam optimizer \cite{kingma_adam:_2017}. The XceptionNet \cite{chollet_xception:_2017} serves as the base model, acting as the encoder. The output of the training process is a set of two distances, which are input into the Triplet Loss function \cite{hoffer_deep_2018}, as shown in Equation \ref{EQN:triplet_loss}: 
\begin{equation}
\begin{split}
Triple Loss = \sum_{N}^{i} [\|{f(x_i^a)- f(x_i^p)}\|^2 
\\ - \|{f(x_i^a)- f(x_i^n)}\|^2 + \alpha]
\label{EQN:triplet_loss}
\end{split}
\end{equation}
Figure \ref{FIG:siamese_working} illustrates the architectural functioning of the SimplyMime's authentication module.

In the Equation \ref{EQN:triplet_loss}, the function $f(x)$ maps each input image to a 4096-dimensional embedding, represented by a tensor. The input images, denoted as $a$, $p$, and $n$, respectively correspond to the anchor, positive, and negative samples used in the triplet loss function. The margin parameter $\alpha$ is used to control the relative distance between the positive and negative pairs, with the goal of maximizing the separation between them.

\subsection{Hardware Implementation and Setup}

\begin{figure*} [htbp]
\centering
	\includegraphics[width=\textwidth]{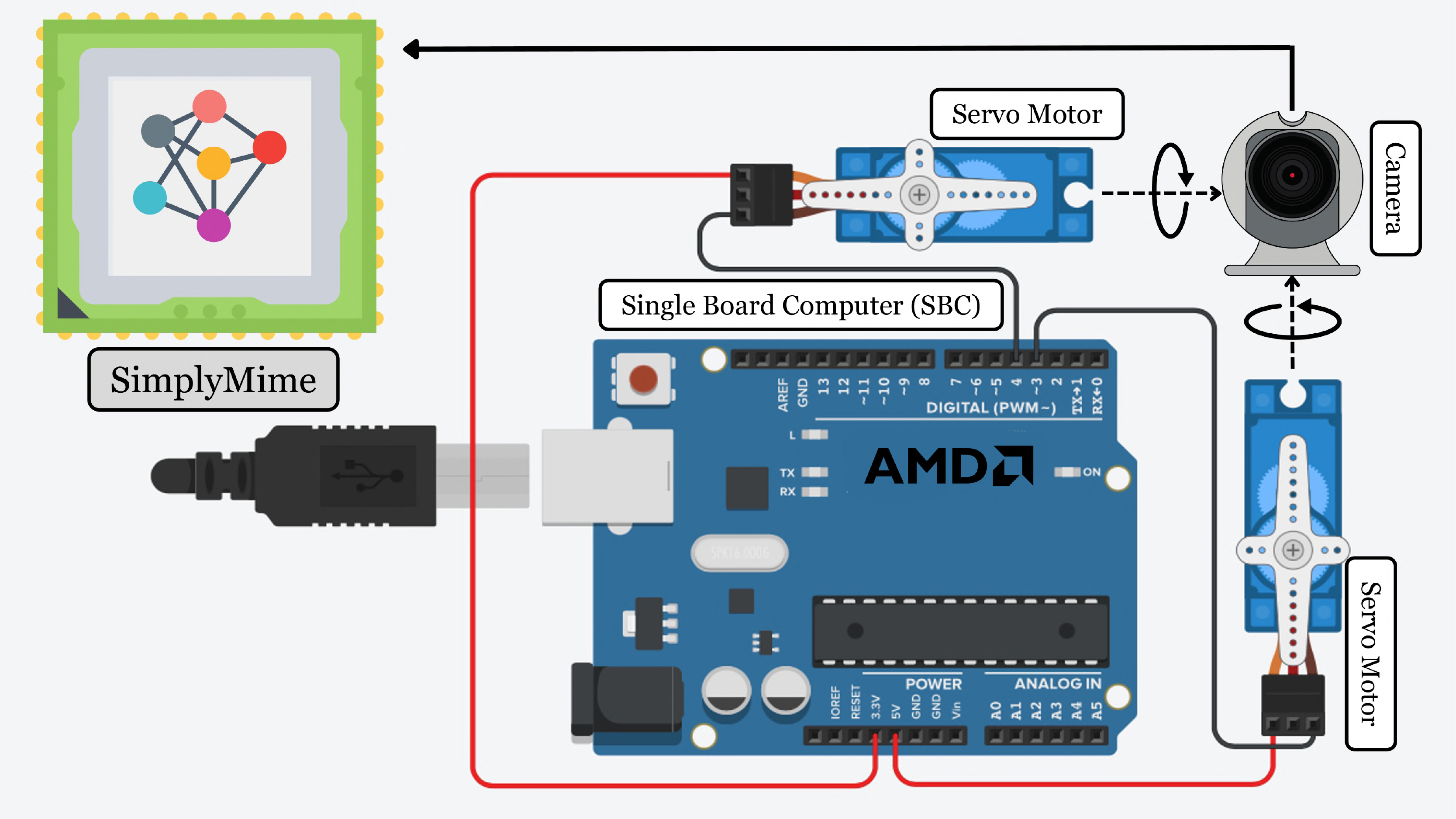}
\caption{Circuit implementation of SimplyMime}
\label{FIG:circuit}
\end{figure*}

\begin{figure} [htbp]
\centering
	\includegraphics[width=0.65\textwidth]{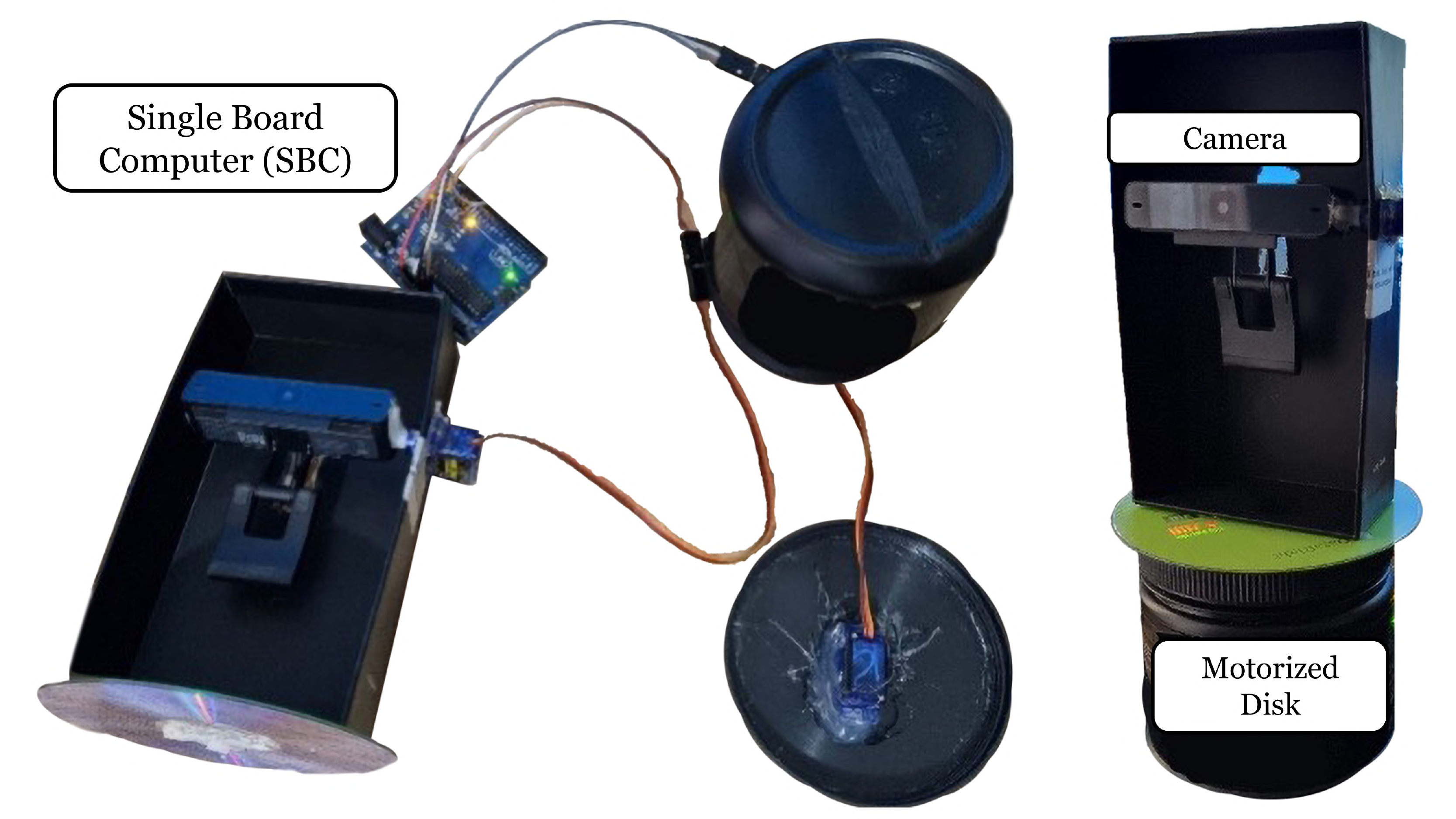}
\caption{Hardware setup of SimplyMime}
\label{FIG:hardware_setup}
\end{figure}

The hardware setup of SimplyMime is a crucial aspect of its overall design and functionality. The camera is mounted on a cuboidal cardboard structure that is equipped with a motor on its right side, which enables control over the Y-axis movement. The cardboard structure is further mounted on a CD disk which is connected to another motor, allowing for control over the X-axis movement. These motors are connected to an On-board microcontroller, which facilitates communication between the hardware components and the software algorithms. Figure \ref{FIG:hardware_setup} provide a visual representation of the hardware setup, while the circuit diagram in the Figure \ref{FIG:circuit} illustrates the connections between the various components. The hardware setup is designed to be compact, portable, and easy to set up, enabling users to easily control consumer electronics with hand gestures.

\section{Experimental Results}
\label{Sec:Experimental_validation}

\begin{figure} [htbp]
\centering
\subfigure[HANDS Dataset \cite{nuzzi_hands:_2021}]{
	\includegraphics[width=0.75\textwidth]{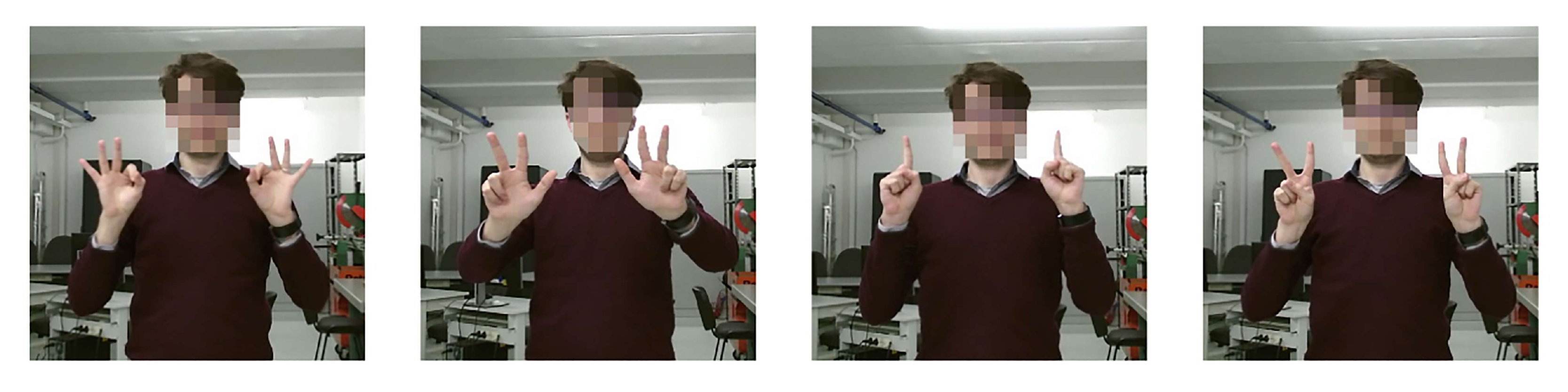}
   	\label{FIG:HANDS}
}
\subfigure[FreiHand Dataset \cite{zimmermann_freihand:_2019}]{
	
	\includegraphics[width=0.75\textwidth]{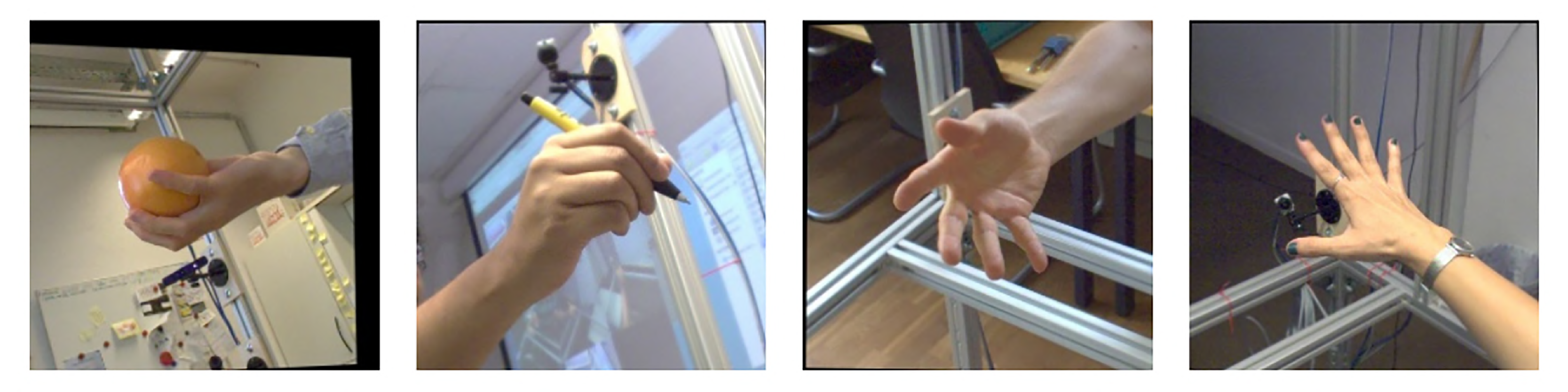}
	\label{FIG:FREIHAND}
}
\subfigure[CASIA Dataset \cite{sun_ordinal_2005}]{
	
	\includegraphics[width=0.75\textwidth]{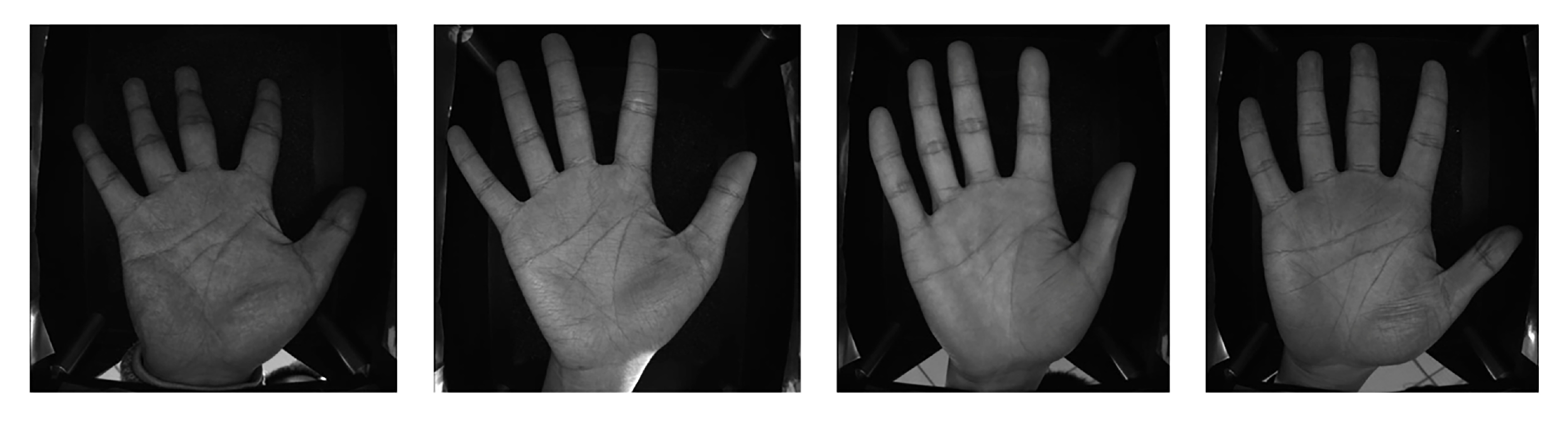}
	\label{FIG:CASIA}
}

\caption{Benchmark datasets utilized in SimplyMime Experimentation}
\end{figure}

In order to thoroughly evaluate the performance and accuracy of our hand gesture recognition model, we conducted extensive testing on two benchmark datasets. The first dataset was utilized to assess the model's ability to accurately detect gestures in images, resulting in an impressive 96.16\% accuracy \cite{nuzzi_hands:_2021}. The second dataset \cite{zimmermann_freihand:_2019} was utilized to evaluate the model's detection rate, yielding a accuracy of 87.37\%. Additionally, we also evaluated our Siamese network-based palmprint authentication system using the CASIA dataset \cite{sun_ordinal_2005}, resulting in an accuracy rate of over 90\%. These results demonstrate the effectiveness and robustness of our proposed model in identifying and authenticating hand gestures for controlling consumer electronics. Figures \ref{FIG:HANDS}, \ref{FIG:FREIHAND}, and \ref{FIG:CASIA} set our few samples from all the utilised benchmark datasets. 

\subsection{HANDS Dataset}

The HANDS dataset comprises a substantial corpus of samples collected from 5 diverse participants, consisting of 3 male and 2 female individuals \cite{nuzzi_hands:_2021}. The participants were instructed to perform 16 pre-defined gestures, comprising of 12 single-handed gestures performed with both arms, and 4 double-handed gestures. Each gesture was captured in 150 RGB frames, resulting in a total of 11,250 images across all participants. Additionally, the gestures were captured from varying distances, resulting in a diverse range of depths and variations in the hand poses.

\begin{figure*} [htbp]
\centering
	\includegraphics[width=0.85\textwidth]{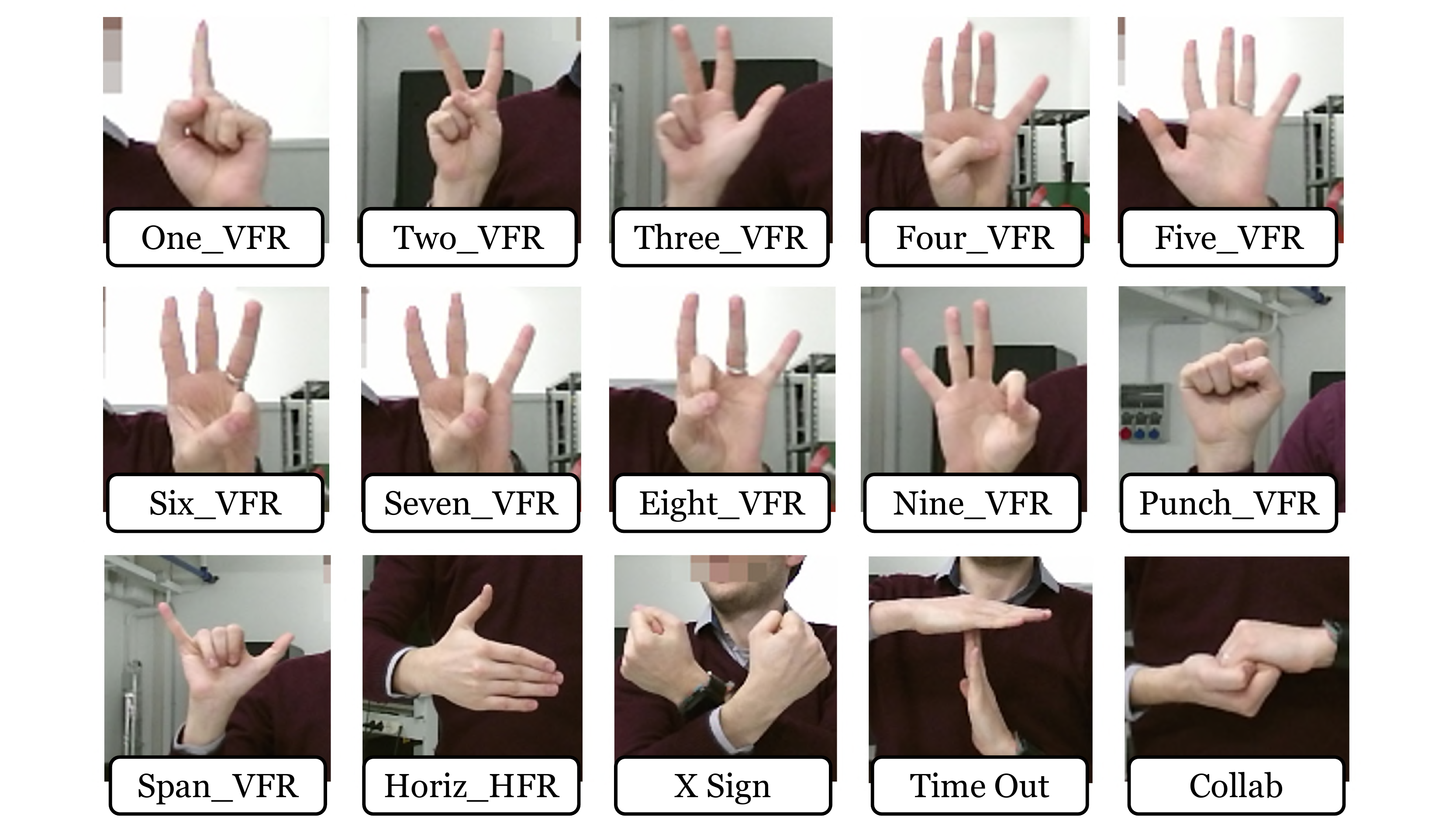}
\caption{Results of various Gestures classified and labeled}
\label{FIG:hands_gestures}
\end{figure*}

Upon evaluation of the dataset, our proposed model achieved an overall accuracy of 96.16\%. The evaluation metric, the recall in particular, is extremely crucial for evaluating SimplyMime as they provide an understanding of how well the model is able to correctly identify hand gestures among the input data, and balance the trade-off between false positives and false negatives. These metrics aid in evaluating the overall performance of the model and the ability of the model to generalize to unseen data. The Recall is used compute the proportion of true positive predictions among all actual positive instances in the data. In the context of SimplyMime, a high recall score indicates that the model is able to identify a high proportion of true hand gestures among all gestures present in the input data. These performances are further highlighted in Figure \ref{FIG:hands_gestures} and Table \ref{tab:handrec_results}, where the gesture-specific results are also illustrated.

\begin{table*}[t]
\centering
\caption{Results of our Hand Gesture Recognition Model on the HANDS dataset \cite{nuzzi_hands:_2021}}
\label{tab:handrec_results}
\resizebox{\textwidth}{!}{%
\begin{tabular}{ccccccc}
\hline 
\textbf{Gesture name} & \textbf{Total frames} & \textbf{Accurately predicted frames} & \textbf{Falsely predicted frames} & \textbf{Accuracy \%} & \textbf{Error \%} & \textbf{Recall} \\
\hline \hline
Collab & 750 & 670 & 80 & 89.33 & 10.67 & 0.89 \\
TimeOut & 750 & 686 & 64 & 91.46 & 8.53 & 0.91 \\
XSign & 750 & 704 & 46 & 93.86 & 6.13 & 0.94 \\
Eight\_VRF & 750 & 708 & 42 & 94.40 & 5.60 & 0.94 \\
Seven\_VRF & 750 & 714 & 36 & 95.20 & 4.80 & 0.95 \\
Eight\_VRF & 750 & 718 & 32 & 95.73 & 4.26 & 0.96 \\
Horiz\_HRF & 750 & 727 & 23 & 96.93 & 3.06 & 0.97 \\
Span\_VRF & 750 & 728 & 22 & 97.06 & 2.93 & 0.97 \\
Six\_VRF & 750 & 732 & 18 & 97.60 & 2.40 & 0.98 \\
Five\_VRF & 750 & 733 & 17 & 97.73 & 2.26 & 0.98 \\
Four\_VRF & 750 & 736 & 14 & 98.13 & 1.86 & 0.98 \\
Three\_VRF & 750 & 738 & 12 & 98.40 & 1.60 & 0.98 \\
Two\_VRF & 750 & 739 & 11 & 98.53 & 1.46 & 0.99 \\
One\_VRF & 750 & 741 & 9 & 98.80 & 1.20 & 0.99 \\
Punch\_VRF & 750 & 742 & 8 & 98.93 & 1.06 & 0.99 \\
\hline
\textbf{Total} & \textbf{11250} & \textbf{10816} & \textbf{434} & \textbf{96.16} & \textbf{3.85} & \textbf{0.9613} \\
\hline
\end{tabular}%
}
\end{table*}

Furthermore, in order to evaluate the effectiveness of our proposed system, we conducted a comparative analysis against other notable models in the field. One such study utilized Generative Adversarial Networks for hand gesture detection \cite{feng_gesture_2022}, while another employed a similar pipeline to ours and leveraged the popular MobileNetV2 architecture as the baseline model \cite{dang_improved_2022}. Despite the high accuracy results achieved by these models, they required significant computational power. In contrast, our proposed system, SimplyMime, achieved comparable performance and precision while requiring significantly less computational resources as shown in Table \ref{tab:handrec_comparision}. 
 
\begin{table}[htbp]
\centering
\caption{Comparison of our Hand Gesture Recognition model against existing solution}
\label{tab:handrec_comparision}
\begin{tabular}{ p{4cm} p{3.3cm} p{1.6cm} }
 \hline
 Research & Architecture Used & Accuracy \\ 
 \hline \hline 
 Feng et al. (2022) \cite{feng_gesture_2022} & GANs & 96\% \\  
 Dang et al. (2022) \cite{dang_improved_2022} & MobileNetV2 & 94\%  \\  
 \textbf{SimplyMime} & CNN based skeletal pose estimation & 96\% \\  
\hline
\end{tabular}
\end{table}

\begin{figure*} [!htbp]
\centering
	\includegraphics[width=\textwidth]{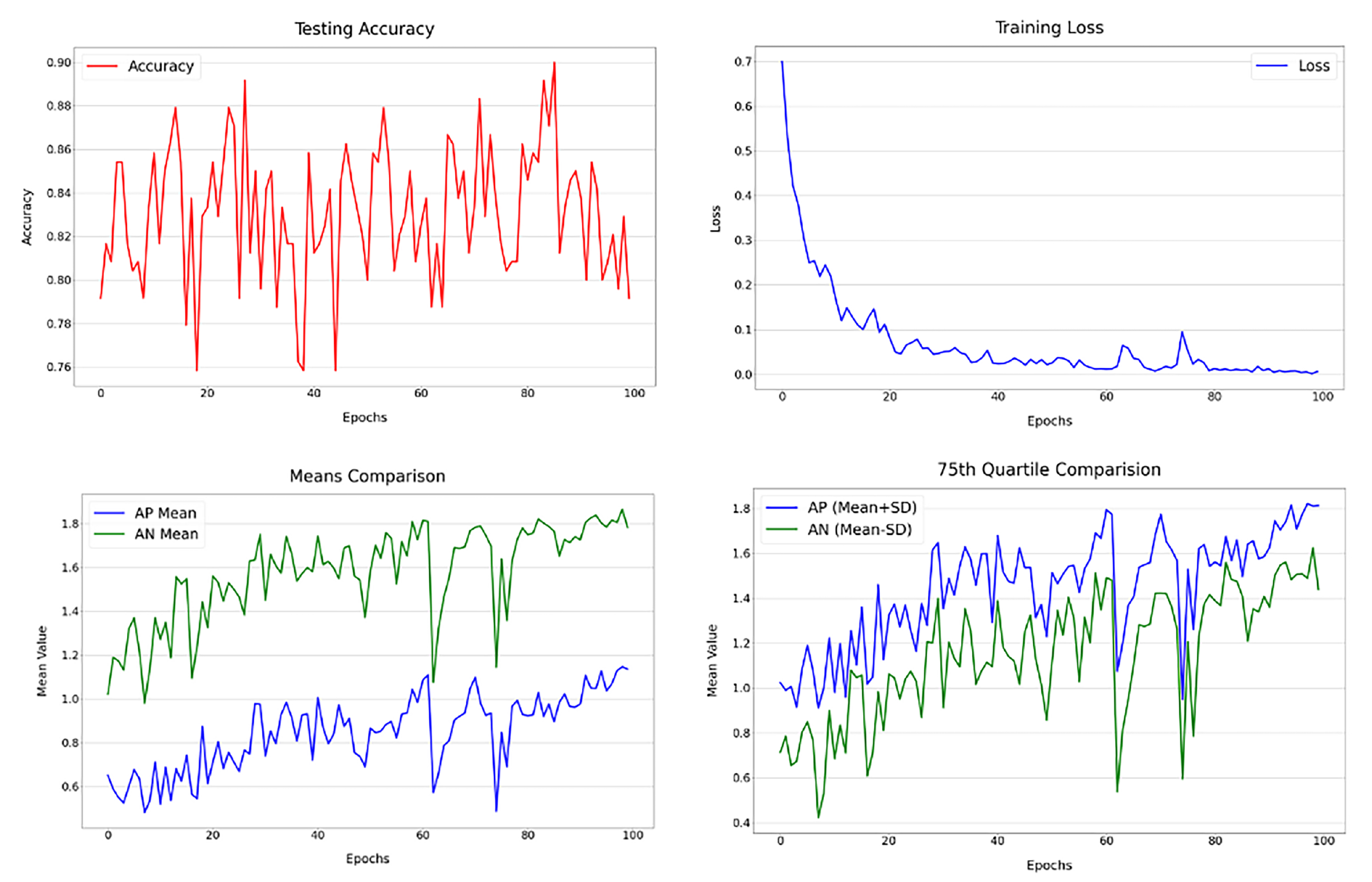}
\caption{Performance measures of the proposed palm print authentication model}
\label{FIG:Siamese_NN_Results}
\end{figure*}

\subsection{FreiHand Dataset}

A robust hand gesture-based control system requires a strong hand detector as its backbone. The hand detector's primary function is to accurately localize the hand region within an image, which is crucial for the subsequent gesture recognition stage. Therefore, to evaluate the performance of our hand gesture recognition model, SimplyMime, we also conducted evaluations to measure the capacity of our hand detector. The results of these evaluations provide insight into the model's ability to accurately detect and localize the hand region, which is crucial for the overall performance of the system. Additionally, by comparing the results of our hand detector with those of other models, we can gain a better understanding of the performance of our system in relation to the state-of-the-art.

The FreiHand dataset is a benchmark dataset specifically designed to evaluate the performance of hand detection and hand pose estimation models \cite{zimmermann_freihand:_2019}. The dataset contains over 32,560 frames of synchronized RGB and depth data, captured using Microsoft Kinect v2 sensors, collected by 32 people. This dataset is considered to be one of the most challenging datasets for hand detection and pose estimation, as it contains a wide range of hand poses and motion, captured under various lighting conditions and backgrounds. To evaluate the performance of SimplyMime, we used the Freihand dataset to test our model's ability to detect hands and estimate hand poses. The dataset was particularly useful in evaluating the model's performance under challenging conditions such as low resolution, low contrast, and occlusions \cite{zimmermann_freihand:_2019}. Our model was able to achieve a high level of accuracy in detecting hands and estimating hand poses, even under these challenging conditions. Our system's results on the dataset are depicted in Table \ref{tab:friehand_results}.

\begin{table}[htbp]
\centering
\caption{Model's performance on FrieHand dataset \cite{zimmermann_freihand:_2019}}
\label{tab:friehand_results}
\begin{tabular}{ p{6cm} p{2cm}}
 \hline
 \textbf{Evaluation Metric} & Value \\ 
 \hline \hline
 \textbf{Total images} & 32560\\
 \textbf{Truly detected images} & 28448\\
 \textbf{Falsely detected images} & 4112\\
 \textbf{Accuracy} & 87.37\% \\
 \textbf{Error} & 12.62\%\\
 \hline
\end{tabular}
\end{table}

\begin{figure} [!htbp]
\centering
	\includegraphics[width=0.7\textwidth]{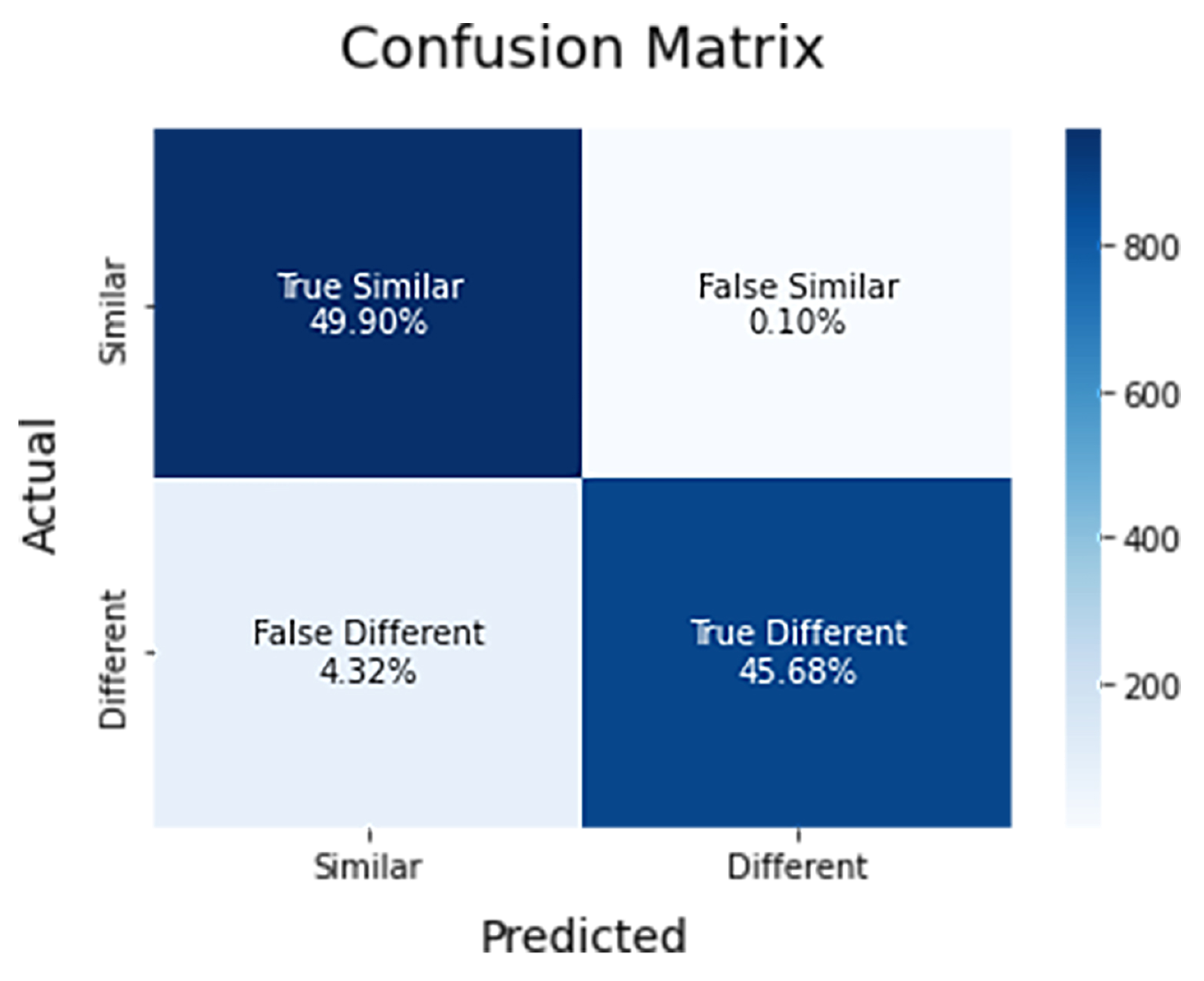}
\caption{Confusion matrix generated by the Siamese Networks}
\label{FIG:conf_mat}
\end{figure}

\subsection{Evaluation of Palm Print Identification Module}

In addition to evaluating the performance of our hand gesture recognition model, we also sought to assess the accuracy of our palmprint identification component. To do so, we utilized the CASIA Dataset \cite{sun_ordinal_2005}, which comprises a large and diverse collection of palmprint images. The dataset includes images from over 100 individuals, captured under various lighting conditions. However, we utilised a subset of the data, particularly the images taken in lowest wavelength sample and white light specifically. As a result, we acquired a dataset with 12000 sample in total. Further on, we pre-processed the dataset leveraging a custom data loader that generated test triplets from the CASIA dataset, where a triplet consists of an anchor image, a positive image, and a negative image. By training our model on these triplets, it is able to learn to differentiate between the palmprints of different individuals and accurately identify a user based on their palmprint. We trained our model using a triplet loss function \cite{muller_few-shot_2022} and Adam optimizer \cite{kingma_adam:_2017}, which allows the model to optimize its error by comparing the similarity between the anchor and positive images to that of the anchor and negative images. This approach allows the model to learn the underlying features of a palmprint that are unique to an individual, which enables it to accurately identify a user based on their palmprint. The proposed model's training metrics are depicted in the Figure \ref{FIG:Siamese_NN_Results}.

One of the key factors for optimized performance of our model is the Triplet Loss \cite{muller_few-shot_2022}. We utilised Triplet loss to compare the relative similarity of three inputs: an anchor image, a positive image, and a negative image. The anchor image is taken to be the "neutral" image, while the positive image is a image from the same subject, in our case, and the negative image is from a different subject. The functions was utilised to minimize the distance between the anchor and positive images, while maximizing the distance between the anchor and negative images. This is done by adjusting the model's weights and biases to better discriminate between the three inputs. As a result, the model becomes better at recognizing the similarities and differences between the anchor and positive images, and can more effectively differentiate between the anchor and negative images. The Figure \ref{FIG:conf_mat} illustrates the confusion matrix acquired from the test set of the data.

Finally, the evaluation results of our proposed SimplyMime model on multiple benchmark datasets indicate its effectiveness in both hand gesture recognition and palmprint identification. Compared to existing solutions, our model has demonstrated superior performance. Further, by incorporating palmprint identification as an additional security measure, the model's practicality and usability in real-world applications have been further enhanced, positioning it as a viable alternative to conventional remote control devices. The novel combination of hand gesture recognition and palmprint identification in one system presents an exciting advancement in the field of human-computer interaction.

\section{Conclusion}
\label{Sec:Conclusion}

In conclusion, this paper presents SimplyMime, a novel hand gesture-based control system that aims to provide an immersive, efficient, and secure user experience while eliminating the need for multiple remote controls for consumer electronics. The system leverages advanced hand gesture recognition techniques, incorporating the latest developments in Artificial Intelligence and Human-Computer Interaction, to create a sophisticated architecture that can recognize a wide range of hand gestures with exceptional accuracy. Additionally, SimplyMime incorporates a palm print authentication module, which enhances the security of the system by ensuring that only authorized users can access the device. Through thorough testing and evaluation, SimplyMime demonstrated remarkable performance, achieving high accuracy levels of 96.16\%, 87.37\%, and 90\% in hand detection, recognition, and palm print authentication, respectively. These results served as a testament to the effectiveness and efficiency of SimplyMime. Overall, SimplyMime offers significant advantages over traditional remote control systems, making it an excellent alternative for users looking for a more intuitive and efficient way of controlling their consumer electronics.

Despite the impressive performance of SimplyMime, there is still scope for improvement and further research. In the future, we plan to enhance the robustness of the system by incorporating additional sensors, such as proximity and depth sensors, to improve the accuracy and reliability of the system. Moreover, we aim to reduce the computational power required while improving the accuracy of the model. Another area of future research is the potential for SimplyMime to be integrated into other applications, such as virtual and augmented reality, to expand its capabilities and utility. By incorporating these enhancements, SimplyMime has the potential to become a highly versatile and widely adopted hand gesture-based control system that can revolutionize the way we interact with consumer electronics.


\bibliographystyle{IEEEtran} 
\bibliography{Bibliography_SimplyMime}

\begin{IEEEbiography}
[{\includegraphics[height=1.2in,keepaspectratio]{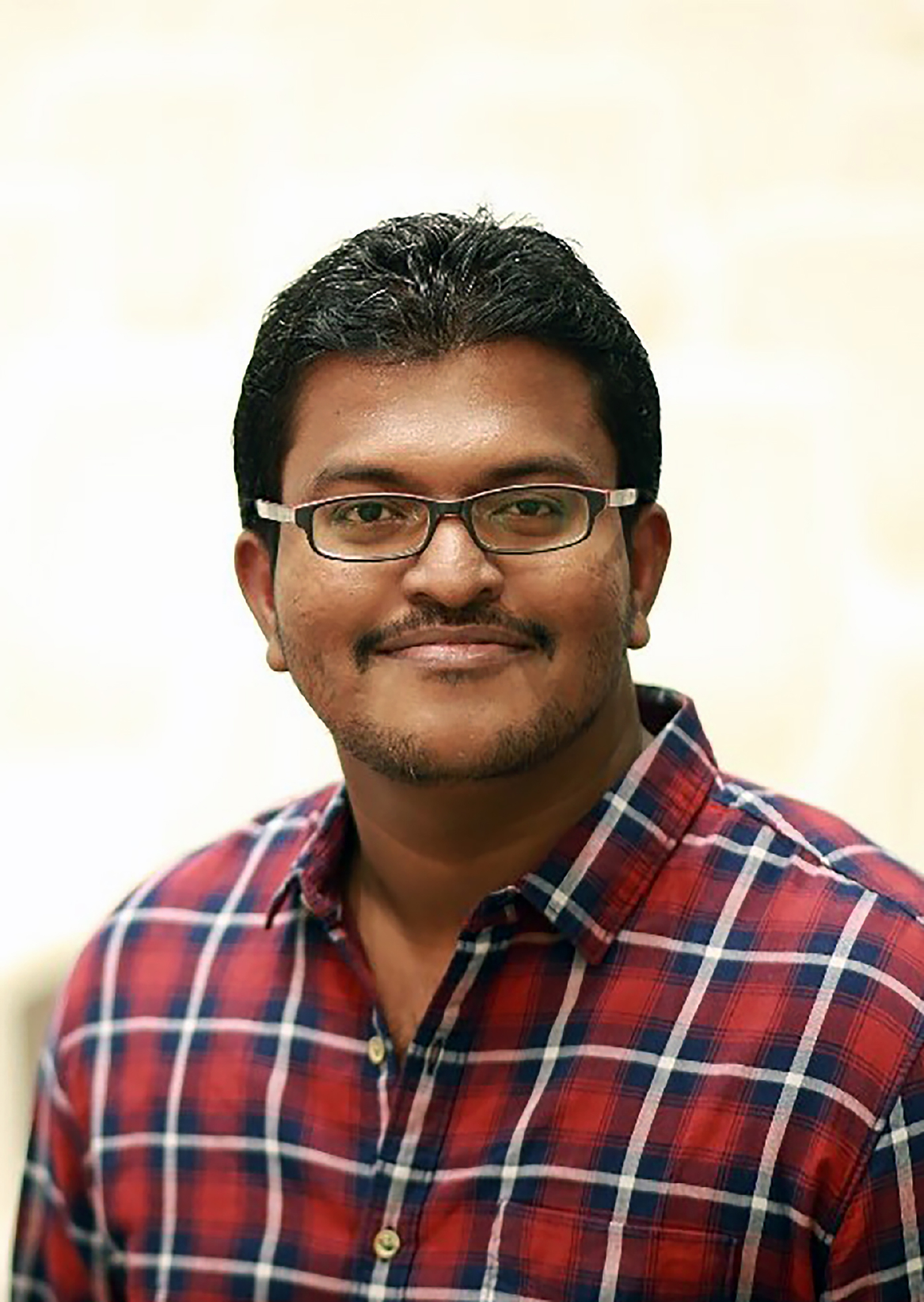}}]
{Sibi C. Sethuraman} (M'18) received his Ph.D from Anna University in the year 2018. He is an Associate Professor in the School of Computer Science and Engineering at Vellore Institute of Technology – Andhra Pradesh (VIT-AP) University. 
Further, he is the coordinator for Artificial Intelligence and Robotics (AIR) Center at VIT-AP.
He is an active reviewer in many reputed journals of IEEE, Springer, and Elsevier. He is a recipient of DST fellowship.
\end{IEEEbiography}


\begin{IEEEbiography}
[{\includegraphics[height=1.2in,clip,keepaspectratio]{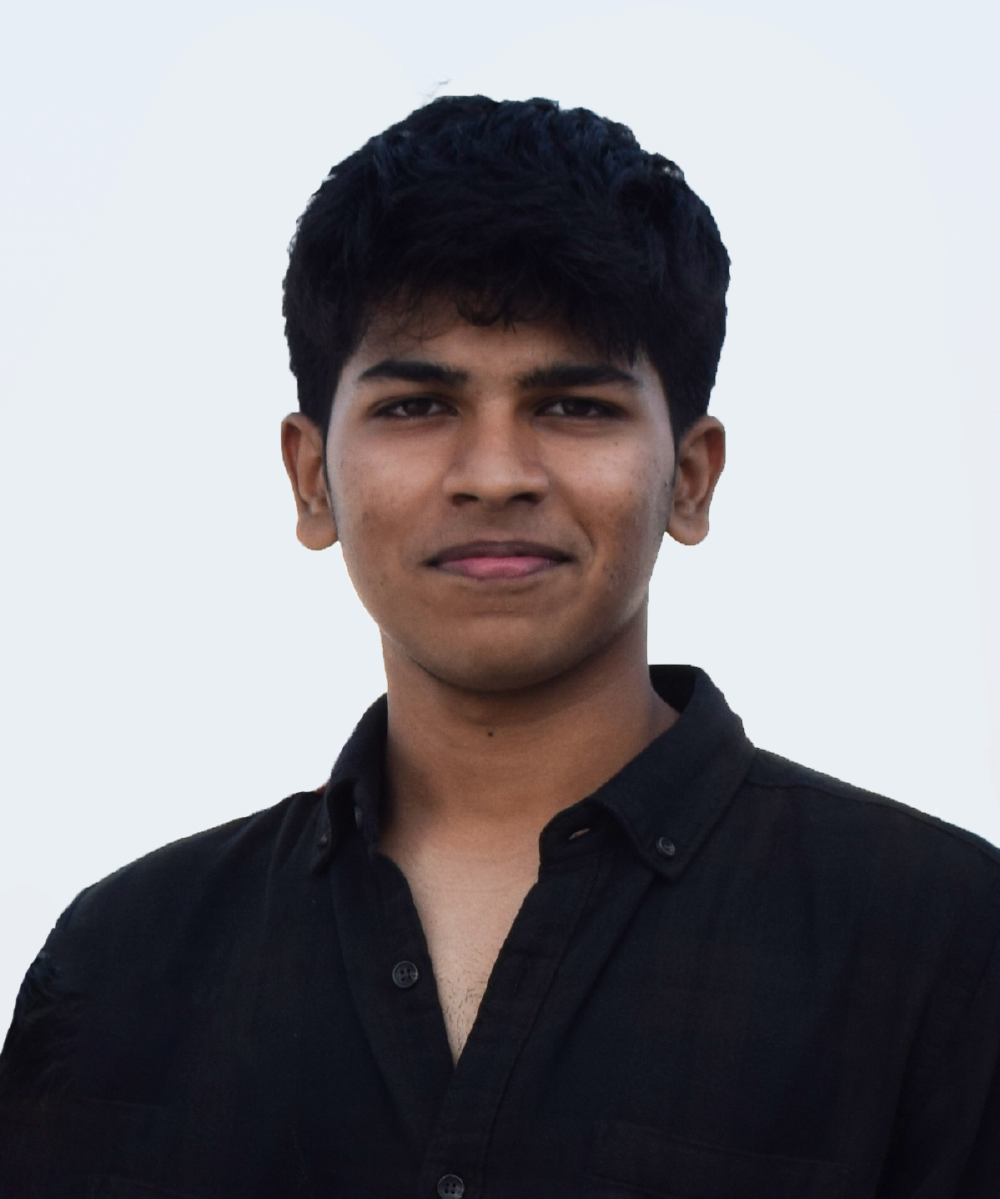}}]
{Gaurav Reddy Tadkapally} (S'19) is a Bachelor of Technology student at Vellore Institute of Technology, Amaravati (VIT-AP). He specializes in Deep Learning and the applications of Artificial Intelligence in Consumer Electronics and Embedded Hardware. As a member of the Artificial Intelligence and Robotics center at VIT-AP, Gaurav has developed several cutting-edge technologies, including iDrone, an IoT-powered drone for detecting wildfires, and MagicEye, an intelligent wearable designed towards
independent living of Visually impaired.
\end{IEEEbiography}

\begin{IEEEbiography}
[{\includegraphics[height=1.2in,clip,keepaspectratio]{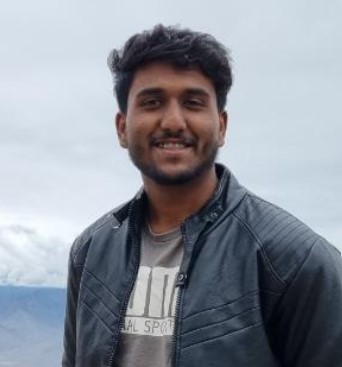}}]
{Athresh Kiran} (M'22) received the bachelor's degree in Computer Science \& Engg. from the VIT-AP University, India, in 2021. He is currently working in Parallel Reality. His research interest includes reality platforms, Metaverse, IT, full stack development etc. 
\end{IEEEbiography}

\begin{IEEEbiography}
[{\includegraphics[height=1.25in, keepaspectratio]{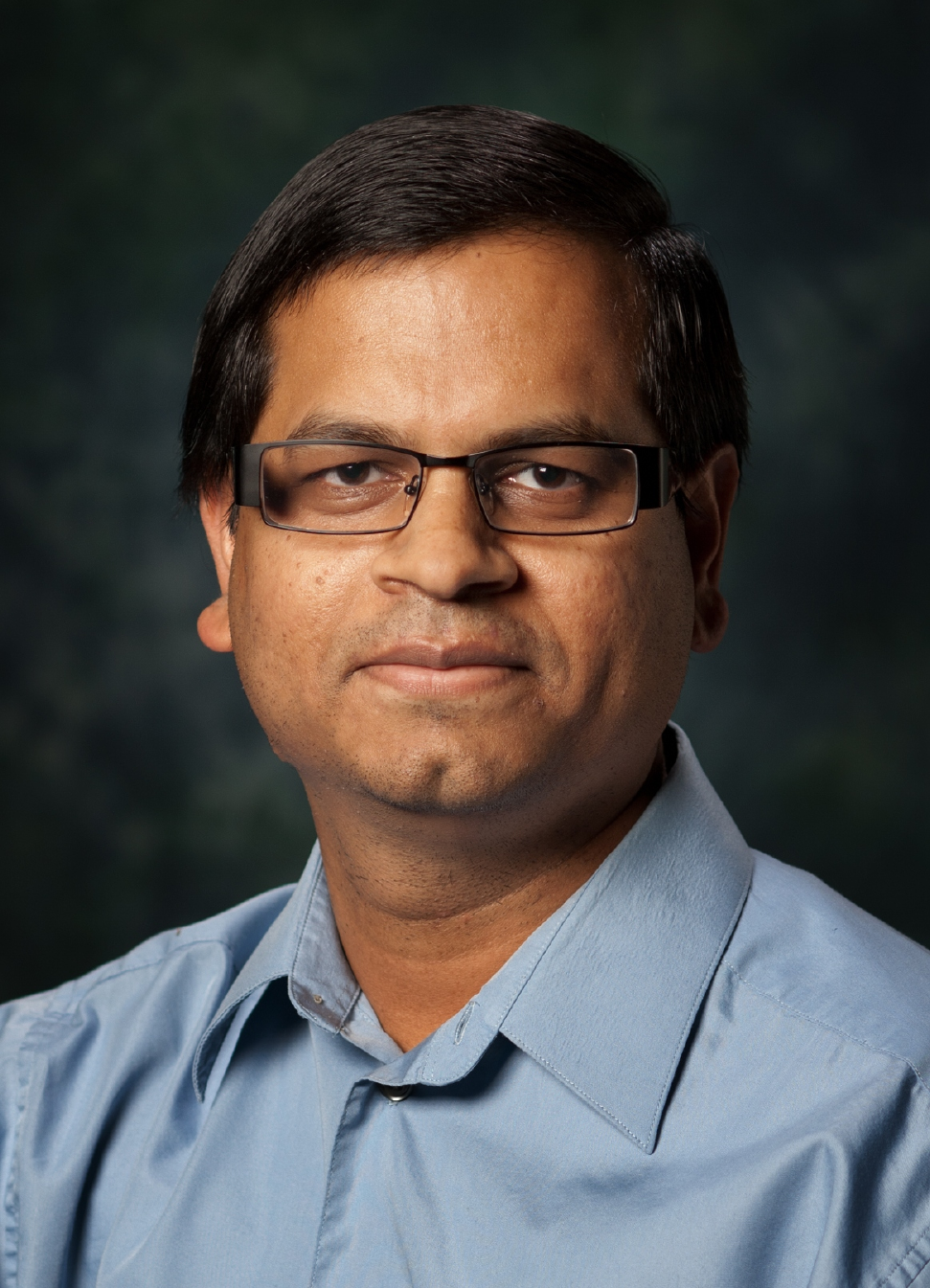}}]
{Saraju P. Mohanty} (Senior Member, IEEE) received the bachelor's degree (Honors) in electrical engineering from the Orissa University of Agriculture and Technology, Bhubaneswar, in 1995, the master's degree in Systems Science and Automation from the Indian Institute of Science, Bengaluru, in 1999, and the Ph.D. degree in Computer Science and Engineering from the University of South Florida, Tampa, in 2003. He is a Professor with the University of North Texas. His research is in "Smart Electronic Systems" which has been funded by National Science Foundations (NSF), Semiconductor Research Corporation (SRC), U.S. Air Force, IUSSTF, and Mission Innovation. He has authored 450 research articles, 5 books, and 9 granted and pending patents. His Google Scholar h-index is 49 and i10-index is 211 with 10,800 citations. He is regarded as a visionary researcher on Smart Cities technology in which his research deals with security and energy aware, and AI/ML-integrated smart components. He introduced the Secure Digital Camera (SDC) in 2004 with built-in security features designed using Hardware Assisted Security (HAS) or Security by Design (SbD) principle. He is widely credited as the designer for the first digital watermarking chip in 2004 and first the low-power digital watermarking chip in 2006. He is a recipient of 16 best paper awards, Fulbright Specialist Award in 2020, IEEE Consumer Electronics Society Outstanding Service Award in 2020, the IEEE-CS-TCVLSI Distinguished Leadership Award in 2018, and the PROSE Award for Best Textbook in Physical Sciences and Mathematics category in 2016. He has delivered 15 keynotes and served on 14 panels at various International Conferences. He has been serving on the editorial board of several peer-reviewed international transactions/journals, including IEEE Transactions on Big Data (TBD), IEEE Transactions on Computer-Aided Design of Integrated Circuits and Systems (TCAD), IEEE Transactions on Consumer Electronics (TCE), and ACM Journal on Emerging Technologies in Computing Systems (JETC). He has been the Editor-in-Chief (EiC) of the IEEE Consumer Electronics Magazine (MCE) during 2016-2021. He served as the Chair of Technical Committee on Very Large Scale Integration (TCVLSI), IEEE Computer Society (IEEE-CS) during 2014-2018 and on the Board of Governors of the IEEE Consumer Electronics Society during 2019-2021. He serves on the steering, organizing, and program committees of several international conferences. He is the steering committee chair/vice-chair for the IEEE International Symposium on Smart Electronic Systems (IEEE-iSES), the IEEE-CS Symposium on VLSI (ISVLSI), and the OITS International Conference on Information Technology (OCIT). He has mentored 2 post-doctoral researchers, and supervised 14 Ph.D. dissertations, 26 M.S. theses, and 18 undergraduate projects.
\end{IEEEbiography}

\begin{IEEEbiography}
[{\includegraphics[height=1.2in,clip,keepaspectratio]{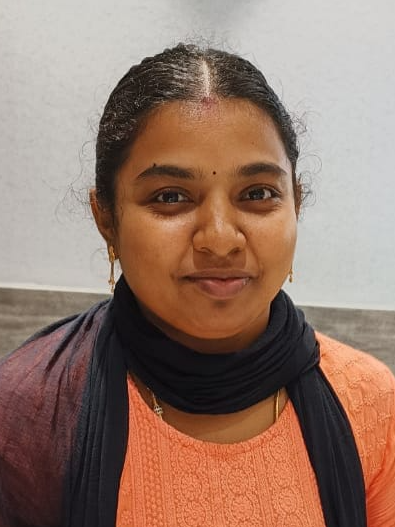}}]
{Anitha S} (M'22) received the bachelor's degree in Electronics \& Communication Engg. from the Anna University, India, in 2017, the master’s degree in Power Electronics from the Anna University University in 2019, and she is currently pursuing her Ph.D. degree in Electronics Engineering from VIT-AP University.
\end{IEEEbiography}

\end{document}